%% file: main.tex
\documentclass[superscriptaddress,twocolumn,10pt,prx,floatfix,nofootinbib]{revtex4-2}
\usepackage[T1]{fontenc}
\usepackage[utf8]{inputenc}
\usepackage{dcolumn}

\usepackage{amsmath, esvect, bm, nth, mathtools, cancel, xfrac}  
\usepackage{amsfonts, amssymb, latexsym, wasysym, xcolor, empheq} 
\usepackage{graphicx,subcaption,caption} 
\usepackage{physics, siunitx}  

\usepackage{bibentry} 

\usepackage{blindtext}
\usepackage{hyperref}

\DeclareMathOperator{\tc}{\textit{t}_\mathrm{c}}
\DeclareMathOperator{\ttc}{2\textit{t}_\mathrm{c}}
\DeclareMathOperator{\Ez}{\overline{\textit{E}_\mathrm{Z}}}
\DeclareMathOperator{\dEz}{\delta\textit{E}_\mathrm{Z}}
\DeclareMathOperator{\VJ}{\textit{V}_\mathrm{J_1}}
\DeclareMathOperator{\tsd}{\textit{t}_\mathrm{sd}}
\DeclareMathOperator{\tsf}{\textit{t}_\mathrm{sf}}

\begin{document}

\title{Interplay of Zeeman Splitting and Tunnel Coupling in Coherent Spin Qubit Shuttling}
\input{author}

\date{July 21, 2025}

\begin{abstract}
Spin shuttling offers a promising approach for developing scalable silicon-based quantum processors
by addressing the connectivity limitations of quantum dots. In this work, we demonstrate high-fidelity bucket-brigade spin shuttling in a silicon MOS device, utilizing Pauli-spin-blockade readout. We achieve an average shuttling fidelity of \SI{99.8}{\percent}. The residual shuttling error is highly sensitive to the ratio between interdot tunnel coupling and Zeeman splitting, with tuning of these parameters enabling up to a 20-fold variation in error rate. An appropriate four-level Hamiltonian model supports our findings. These results provide valuable insights for optimizing high-performance spin-shuttling systems in future quantum architectures.
\end{abstract}

\maketitle

\input{Content}

\appendix
\input{Content_2}

\clearpage

\bibliography{bibliography.bib}

\clearpage

\renewcommand\thefigure{S\arabic{figure}}
\setcounter{figure}{0}  

\onecolumngrid

\input{Supplementary}

\end{document}

%% file: author.tex
\author{Ssu-Chih Lin}
\email{d09222032@ntu.edu.tw}
\affiliation{
    Department of Physics, National Taiwan University, Taipei 106319, Taiwan
}
\affiliation{
    School Electrical Engineering and Telecommunications, University of New South Wales, Sydney, NSW 2052, Australia
}

\author{Paul Steinacker}
\affiliation{
    School Electrical Engineering and Telecommunications, University of New South Wales, Sydney, NSW 2052, Australia
}

\author{MengKe Feng}
\affiliation{
    School Electrical Engineering and Telecommunications, University of New South Wales, Sydney, NSW 2052, Australia
}
\affiliation{
    Diraq Pty. Ltd., Sydney, NSW, Australia
}

\author{Ajit Dash}
\affiliation{
    School Electrical Engineering and Telecommunications, University of New South Wales, Sydney, NSW 2052, Australia
}

\author{Santiago Serrano}
\affiliation{
    School Electrical Engineering and Telecommunications, University of New South Wales, Sydney, NSW 2052, Australia
}
\affiliation{
    Diraq Pty. Ltd., Sydney, NSW, Australia
}

\author{Wee Han Lim}
\affiliation{
    School Electrical Engineering and Telecommunications, University of New South Wales, Sydney, NSW 2052, Australia
}
\affiliation{
    Diraq Pty. Ltd., Sydney, NSW, Australia
}

\author{Kohei M. Itoh}
\affiliation{
    Department of Applied Physics and Physico-Informatics, Keio University, Yokohama 223-8522, Japan
}

\author{Fay E. Hudson}
\affiliation{
    School Electrical Engineering and Telecommunications, University of New South Wales, Sydney, NSW 2052, Australia
}
\affiliation{
    Diraq Pty. Ltd., Sydney, NSW, Australia
}

\author{Tuomo Tanttu}
\affiliation{
    School Electrical Engineering and Telecommunications, University of New South Wales, Sydney, NSW 2052, Australia
}
\affiliation{
    Diraq Pty. Ltd., Sydney, NSW, Australia
}

\author{Andre Saraiva}
\affiliation{
    School Electrical Engineering and Telecommunications, University of New South Wales, Sydney, NSW 2052, Australia
}
\affiliation{
    Diraq Pty. Ltd., Sydney, NSW, Australia
}

\author{Arne Laucht}
\affiliation{
    School Electrical Engineering and Telecommunications, University of New South Wales, Sydney, NSW 2052, Australia
}
\affiliation{
    Diraq Pty. Ltd., Sydney, NSW, Australia
}

\author{Andrew S. Dzurak}
\affiliation{
    School Electrical Engineering and Telecommunications, University of New South Wales, Sydney, NSW 2052, Australia
}
\affiliation{
    Diraq Pty. Ltd., Sydney, NSW, Australia
}

\author{Hsi-Sheng Goan}
\email{goan@phys.ntu.edu.tw}
\affiliation{
    Department of Physics, National Taiwan University, Taipei 106319, Taiwan
}
\affiliation{
    Center for Theoretical Physics, National Taiwan University, Taipei 106319, Taiwan \\
}
\affiliation{
    Center of Quantum Science and Engineering, National Taiwan University, Taipei 106319, Taiwan
}
\affiliation{
   Physics Division, National Center of Theoretical Sciences, Taipei 106319, Taiwan
}

\author{Chih Hwan Yang}
\email{henry.yang@unsw.edu.au}
\affiliation{
    School Electrical Engineering and Telecommunications, University of New South Wales, Sydney, NSW 2052, Australia
}
\affiliation{
    Diraq Pty. Ltd., Sydney, NSW, Australia
}

%% file: Content.tex
\section{Introduction}

Building qubits on silicon nanodevices is an attractive choice due to its compatibility with the established semiconductor industry~\cite{Nat.Electron.4.872-884, npj.Quantum.Inf.10.70, IEEE.10185272, arXiv.2409.03993}. In addition, the electron spin is a natural two-level system and has long coherence times~\cite{Nature.Nanotech.9.981-985}. High fidelity has been demonstrated on various platforms~\cite{Nature.601.343-347, Nature.601.338-342, Sci.Adv.8.eabn5130, Nat.Commun.16.3606}, including aspects such as initialization, manipulation, and readout. Moreover, high-temperature operation~\cite{Nature.580.355-359, Nature.627.772-777} and the cryogenic control interface~\cite{Nat.Electron.4.64-70, Nature.643.382-387} have been successfully established. Recently, devices utilizing the 300-\SI{}{\milli\meter} manufacturing infrastructure have met the requirements of the surface code~\cite{Nature.646.81-87} or have achieved up to 12 qubits~\cite{Nano.Lett.2025.25.793−799}.

Despite the rapid development of silicon quantum dots (QDs), several challenges remain in scaling up. One challenge is the short interaction distance between neighboring spins, which is the basis for the conventional two-qubit gate~\cite{Science309.2180-2184}. Additionally, to enhance the readout and control performance, QDs have to be placed close to these electric gates. These limitations not only restrict the arrangement of the QD array but also reduce qubit connectivity. Furthermore,  this condensed layout of qubits can lead to issues with crosstalk interference~\cite{PhysRevB.110.125414, qgnt-n527}. Besides, as the number of QDs and top gates increases, managing individual voltages through numerous top gates poses significant overhead and interconnect challenges~\cite{Nat.Nanotechnol.19.21-27, xhq3-4jxz}. In addition, the large quantity of cables can introduce excessive heat and noise into the system from room temperature~\cite{Microprocess.Microsyst.67.1-7}.

A promising alternative is to separate the QDs and implement local cryogenic control circuits in between them~\cite{PhysRevApplied.18.024053}. To achieve this protocol, we can couple the qubits through mediators~\cite{Nat.Phys.21.168-174, Adv.Mater.2023.35.2208557} or shuttle spins directly~\cite{Nat.Commun.15.4977, arXiv.2501.17814}. Two primary methods for shuttling are commonly used: bucket-brigade (BB) spin shuttling and conveyor-mode (CV) shuttling. The bucket-brigade method has been demonstrated with high fidelity on different silicon platforms~\cite{PRXQuantum.4.030303, Nat.Commun.12.4114, Nat.Commun.13.5740}. Conveyor-mode shuttling has been proposed as an alternative method~\cite{npj.Quantum.Inf.8.100} and has also yielded high fidelities~\cite{Nat.Commun.15.2296, Nat.Nanotechnol.20.866-872}. Building on these results, there have also been demonstrations of single-qubit control based on spin hopping~\cite{Nat.Commun.16.5605}, alongside proposals for shuttling-based error-correction frameworks~\cite{PRXQuantum.5.040328}.  Additionally, consistent electron shuttling has been successfully shown in foundry QD devices,  proving the potential for applying this protocol in future scalability efforts~\cite{arXiv.2506.15956}.

However, both shuttling protocols exhibit some inadequacies. In the case of CV shuttling, one primary concern is leakage to the excited valley state~\cite{PRXQuantum.5.040322}. This leakage occurs because the spins move continuously within the traveling-wave potential, which may lead them to pass through regions with low valley splitting—especially in materials such as Si/SiGe, where the average valley splitting is generally lower. On the other hand, BB shuttling relies on electrons tunneling through potential barriers between QDs, which limits the shuttling speed to the tunneling rate. Moreover, when the gate voltages are ramped through the anticrossings of the charge state, diabatic Landau-Zener (LZ) transitions can occur~\cite{j.physrep.2010.03.002, Nat.Commun.10.1063, PhysRevB.102.125406, New.J.Phys.20.113029}. In order to mitigate these issues, it is essential that the tunnel coupling $\tc$ is sufficiently high. 

In this work, we investigate how the strength of the tunnel coupling influences spin shuttling. We assess the shuttling fidelity at various levels of tunnel coupling observe mitigation of the dephasing error by a factor of approximately 20. Furthermore, we demonstrate that high fidelities, exceeding \SI{99}{\percent} fidelity, are maintained across a wide range of tunnel coupling under low magnetic field. These experimental results align with the previously proposed theoretical model~\cite{PhysRevB.107.085427}, which accounts for the Zeeman splitting of spin states.

\section{Device Setup}
\begin{figure*}[htp]
	\centering
	\includegraphics[width=1.0\textwidth]{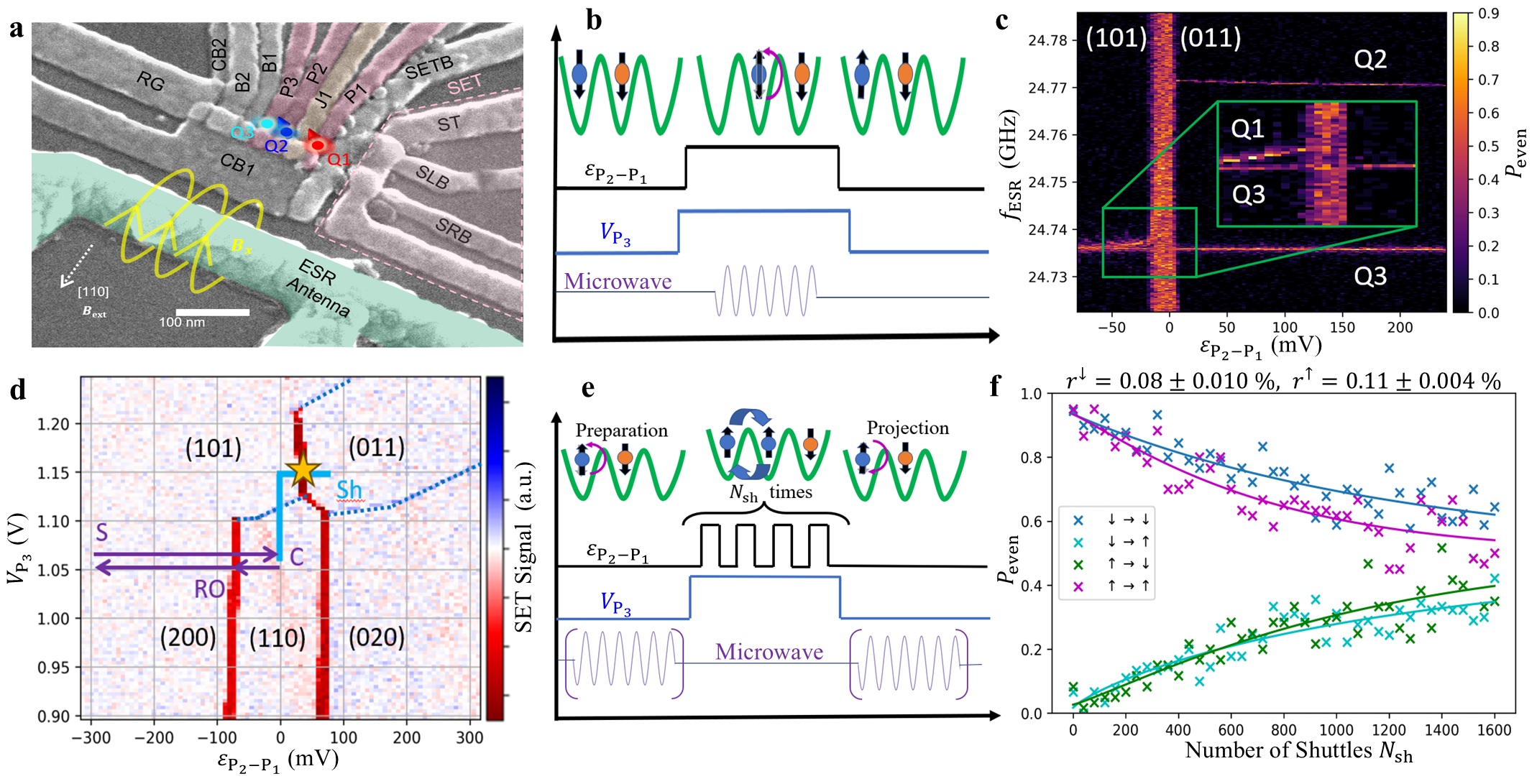}
	\caption{\textbf{The device and polarization shuttling.} 
        \textbf{a,} A false-color scanning-electron-micrograph (SEM) image of a device nominally identical to the one used in this work. The three plunger gates, $\mathrm{P_1}$, $\mathrm{P_2}$ and $\mathrm{P_3}$ are colored in magenta, and the QDs under them are indicated in red, blue, and cyan. The barrier gate $\mathrm{J_1}$ between $\mathrm{Q_1}$ and $\mathrm{Q_2}$ is colored in sienna. The direction of the dc external magnetic field (ac control microwave field) is indicated by white (yellow) arrows. The single-electron transistor (SET) [electron spin resonance (ESR) line] is marked in light pink (green).
        \textbf{b,} A pulse schematic for the pulsed electron spin-orbital spectroscopy (PESOS) map. A constant ESR control pulse is applied after ramping.
        \textbf{c,} The PESOS map near the yellow star in (\textbf{d}) under a 0.89-\SI{}{\tesla} external field. The inset is an enlarged image near 24.735 GHz, where $f_{\mathrm{Q_1}}$ and $f_{\mathrm{Q_3}}$ are located. 
        \textbf{d,} The charge-stability diagram (CSD) as a function of the voltage detuning $\varepsilon_{\mathrm{P_2 \text{-} P_1}}$, which is used in the $\mathrm{Q_1 \text{-} Q_2}$ transition, and the gate voltage $V_\mathrm{P_3}$, which is used in the $\mathrm{Q_2 \text{-} Q_3}$ transition. The current difference at transitions of the $\mathrm{Q_1 \text{-} Q_2}$ is clear in red, while those of the $\mathrm{Q_2 \text{-} Q_3}$ transition are unclear after removing the background and are indicated by the dotted blue lines (see Fig. \ref{fig:CSD}a in the Supplemental Material). The control and readout points are labeled as C and RO, respectively. The cyan path indicates our shuttling protocol from the (110) to the (011) charge state.
        \textbf{e,} The pulse schematic for the polarization-shuttling experiment. Depending on the states prepared or the measurement projections, $\pi$ rotations are applied before or after the consecutive shuttling.
        \textbf{f,} The probabilities of finding spin-up or spin-down postshuttling if spin-up or spin-down are prepared under a 0.17-\SI{}{\tesla} external field. The solid curves are fits to the data (crosses); the shuttling depolarizing rates $r^{\downarrow(\uparrow)}$ and their errors are calculated from the exponential fits in Fig. \ref{fig:eigen} in the Supplemental Material.
    }
	\label{fig:fig1}
\end{figure*}

The device is a silicon MOS QD system on an isotopically enriched $^{28}\mathrm{Si}$ substrate with 800-\SI{}{ppm} residual $^{29}\mathrm{Si}$~\cite{MRS.Communications.4.143-157}. Three layers of aluminum gates define a three-QD array isolated from an adjacent reservoir (Fig.~\ref{fig:fig1}a)~\cite{Appl.Phys.Lett.95.242102}. Three QDs, referred to as $\mathrm{Q_1}$, $\mathrm{Q_2}$, and $\mathrm{Q_3}$, form beneath the plunger gates $\mathrm{P_1}$, $\mathrm{P_2}$, and $\mathrm{P_3}$, respectively. The barrier gate $\mathrm{J_1}$ controls the tunnel coupling $\tc$ between $\mathrm{Q_1}$ and $\mathrm{Q_2}$, which varies exponentially at a rate of \SI{23.7}{dec\per\volt} (Fig.~\ref{fig:fig3}c)~\cite{Nano.Lett.2025.25.10263-10269}. In contrast, we are unable to control the tunnel coupling between $\mathrm{Q_2}$ and $\mathrm{Q_3}$, because the QDs are formed under two neighboring gates that lack an interstitial barrier gate. Besides, the subtle transition between $\mathrm{Q_2}$ and $\mathrm{Q_3}$ in Fig.~\ref{fig:fig1}d suggests a slower tunneling rate, which may account for the poor shuttling fidelity observed between $\mathrm{Q_2}$ and $\mathrm{Q_3}$ in our experiments. This low tunneling rate is attributed to the unintentional formation of $\mathrm{Q_3}$ under $\mathrm{P_3}$, which was originally intended to function as a barrier gate and was situated at a higher gate stack during fabrication. 

We load two electrons into the QDs, and under a static external magnetic field supplied by a superconducting magnet, these two unpaired spins function as qubits. As a result of the poor shuttling between $\mathrm{Q_2}$ and $\mathrm{Q_3}$, we maintain the second spin in the spin-down state and utilize it solely as an ancilla qubit, characterizing only the shuttling across $\mathrm{Q_1}$ and $\mathrm{Q_2}$. A single-electron transistor (SET) and an electron spin resonance (ESR) nearby are utilized for state readout and spin control~\cite{Nature.Nanotech.9.981-985, Nat.Commun.16.3606}. We connect an $LC$ tank circuit to the source of the SET for charge readout in rf reflectometry mode~\cite{Appl.Phys.Lett.92.112103, Appl.Phys.Rev.10.021305}. 

\begin{figure*}
	\centering
	\includegraphics[width=1.0\textwidth]{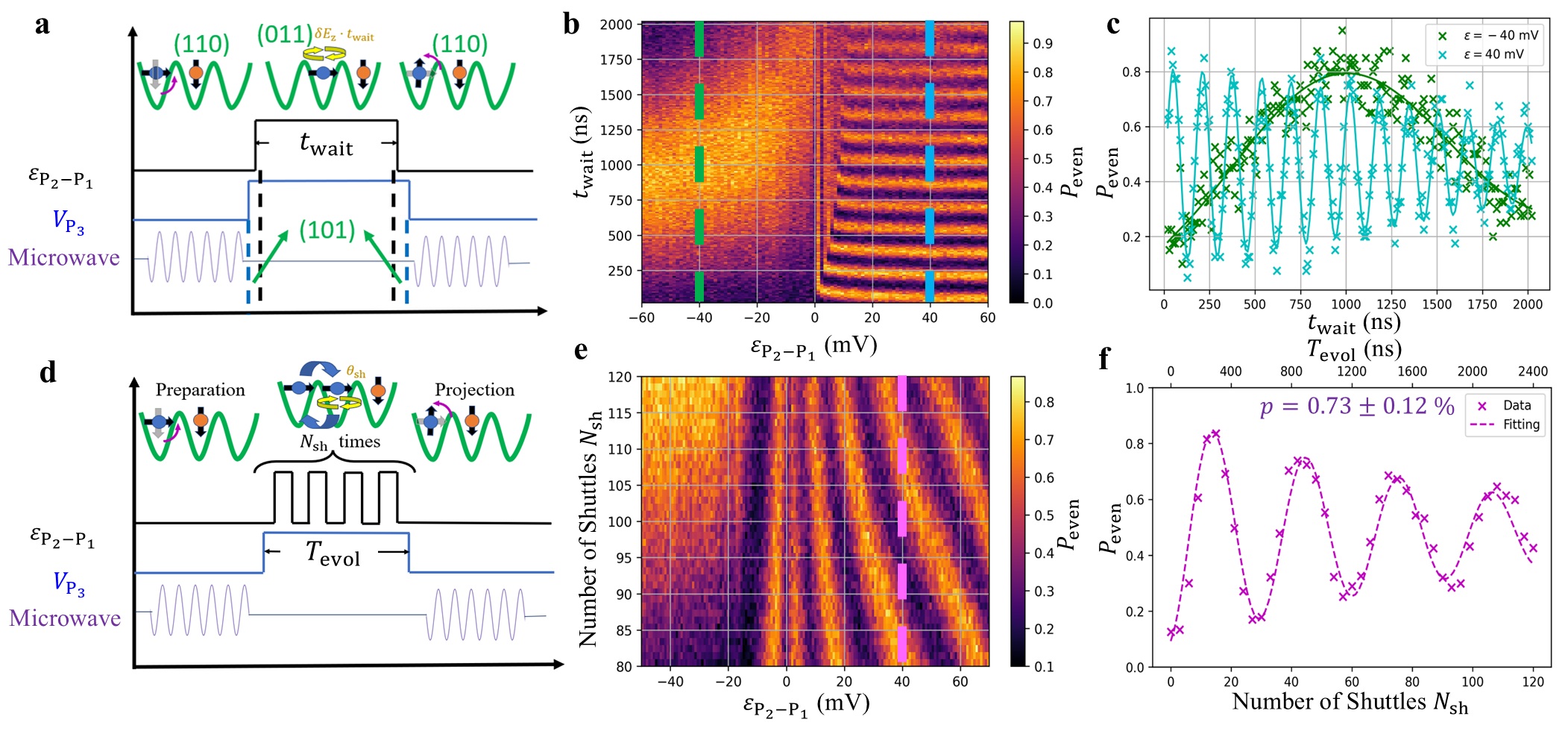}
	\caption{\textbf{Phase-coherent shuttling spectroscopy under an 0.17-\SI{}{\tesla} external field.}
		\textbf{a,} The pulse schematic for the shuttling spectroscopy. The spin in $\mathrm{Q_1}$ is first rotated to the equatorial state at the control point. After ramping the voltage, the spin accumulates a phase during the wait time $t_\mathrm{wait}$ before ramping back to the (110) state. A second $X(\pi/2)$ gate is applied to project the phase onto the polarization in the measurement basis.
        \textbf{b,} Shuttling spectroscopy near the $\mathrm{Q_1 \text{-} Q_2}$ change transition. The continuous fringe evolution demonstrates the phase coherence when shuttling.
        \textbf{c,} A line-cut of (\textbf{b}) at $\varepsilon_\mathrm{P_2 \text{-} P_1} = \pm \SI{40}{\milli\volt}$.
        \textbf{d,} The pulse schematic for the consecutive shuttling spectroscopy. Between two single-qubit gates at the control point, the voltage is ramped back and forth between $\mathrm{Q_1}$ and $\mathrm{Q_2}$ repeatedly in a total evolution time $T_\mathrm{evol}$.
        \textbf{e,} The consecutive-shuttling spectroscopy near the $\mathrm{Q_1 \text{-} Q_2}$ charge transition.
        \textbf{f,} The shuttle characterization at $\varepsilon_{\mathrm{P_2 \text{-} P_1}} = \SI{40}{\milli\volt}$. The stable oscillation period demonstrates the consistency of each shuttling operation.
    }
	\label{fig:fig2}
\end{figure*}

Initialization begins in the singlet ground state after waiting for \SI{300}{\nano\second} at the start point (S) deep in the (200) state, where $(l m n)$ represents the number of electrons in $\mathrm{Q_1}$, $\mathrm{Q_2}$, and $\mathrm{Q_3}$, respectively (Fig.~\ref{fig:fig1}d). We then adiabatically ramp the voltage from the $\ket{\mathrm{S}(200)}$ state to the $\ket{\uparrow, \downarrow, 0}$ state in the (110) charge configuration. The resulting states after ramping are verified from the branching of resonance frequencies in the pulsed electron spin-orbital spectroscopy (PESOS) map~\cite{Nat.Nanotechnol.18.131-136} when two spins exhibit exchange coupling~\cite{Nat.Commun.16.3606}. Following the ramping process, we apply an additional $\pi$ gate on the spin in $\mathrm{Q_1}$ at the control point (C), leading to the final state $\ket{\downarrow, \downarrow, 0}$. We employ the heralded-initialization protocol to confirm the initial state and improve the state-preparation fidelity for the subsequent experiments~\cite{Nature.627.772-777, Nat.Commun.16.3606}. Additionally, we implement automatic parameter feedback to maintain the SET at the sensitive point and keep on resonance with the Larmor frequencies of both qubits~\cite{Nat.Electron.2.151-158, Appl.Phys.Lett.124.114003}.

We perform readout of the spin states using Pauli spin blockade (PSB) at the readout point (RO) at the (200) and (110) interdot charge transition. We set the integration time to \SI{300}{\micro\second} for parity readout~\cite{PRXQuantum.2.010303, npj.Quantum.Inf.10.22}; this method effectively distinguishes between the odd states $\{\ket{\uparrow\downarrow}, \ket{\downarrow\uparrow}\}$ and the even states $\{\ket{\uparrow\uparrow}, \ket{\downarrow\downarrow}\}$. Although using two spins for PSB necessitates a more complex shuttling protocol in our system, the PSB readout mitigates the limitation of large Zeeman splitting, in contrast to the Elzerman readout used in previous studies~\cite{Nat.Commun.12.4114}. Consequently, we can conduct operations and investigate the shuttling process under much lower magnetic field strength.

\section{Spin Shuttling}

Our shuttling protocol involves transferring both spins from the original QDs one at a time to the shuttle point (Sh) in the (011) charge state. We first move the electron from $\mathrm{Q_2}$ to $\mathrm{Q_3}$ in its eigenstate state, followed by a transfer of the electron from $\mathrm{Q_1}$ to $\mathrm{Q_2}$, as depicted by the cyan lines in Fig.~\ref{fig:fig1}d. Note that this is the shuttling process that is being characterized in these set of experiments, and the number of transfers across this transition depends on the specific shuttling protocol that we adopt. After these shuttling processes are completed, we apply the voltage pulses in reverse to return to the (110) charge state before readout.

To confirm spin shuttling, we first analyze the PESOS map (Fig.~\ref{fig:fig1}b and c) sweeping the voltage near the (101)-(011) charge transitions, indicated by the yellow star in Fig.~\ref{fig:fig1}d. We apply a constant ESR control pulse near the transition after ramping from the (110) state. We have chosen the power and the duration of the ESR microwave burst to roughly perform an odd number of single-qubit $\pi$ rotations on both spins. We use the PESOS map at \SI{0.89}{\tesla} to visually distinguish the resonance frequencies of the spins in $\mathrm{Q_1}$ and $\mathrm{Q_3}$, the difference between which is smaller than \SI{5}{\mega\Hz} even under this field strength. The PESOS map under \SI{0.17}{\tesla} is shown in Fig.~\ref{fig:CSD} in the Supplemental Material. We observe in Fig.~\ref{fig:fig1}c that as we transition from negative to positive detuning values across the transition at $\varepsilon_\mathrm{P_2 \text{-} P_1}=0$, we have an abrupt change in the resonance frequencies of the two electrons. We identify this as the electron tunneling from $\mathrm{Q_1}$ to $\mathrm{Q_2}$, which has a Zeeman-splitting difference of about \SI{30}{\mega\Hz}. We identify the smooth changing line as the resonance frequency of the electron in $\mathrm{Q_3}$ which would not have moved across the transition. This abrupt change in resonance frequencies can be attributed to the different Zeeman splittings in each dot, mainly due to the variation of the $g$ factor within the device~\cite{PhysRevB.92.201401, PhysRevX.9.021028}. Additionally, the Stark shift of the shuttled spin changes after crossing the transition, indicating that the charge state has altered. This abrupt change of resonance frequency also suggests that our shuttling involves BB shuttling rather than CV shuttling, where the resonance frequencies change smoothly during the transfer~\cite{Nat.Nanotechnol.20.866-872}.

\begin{figure*}
	\centering
	\includegraphics[width=1.0\textwidth]{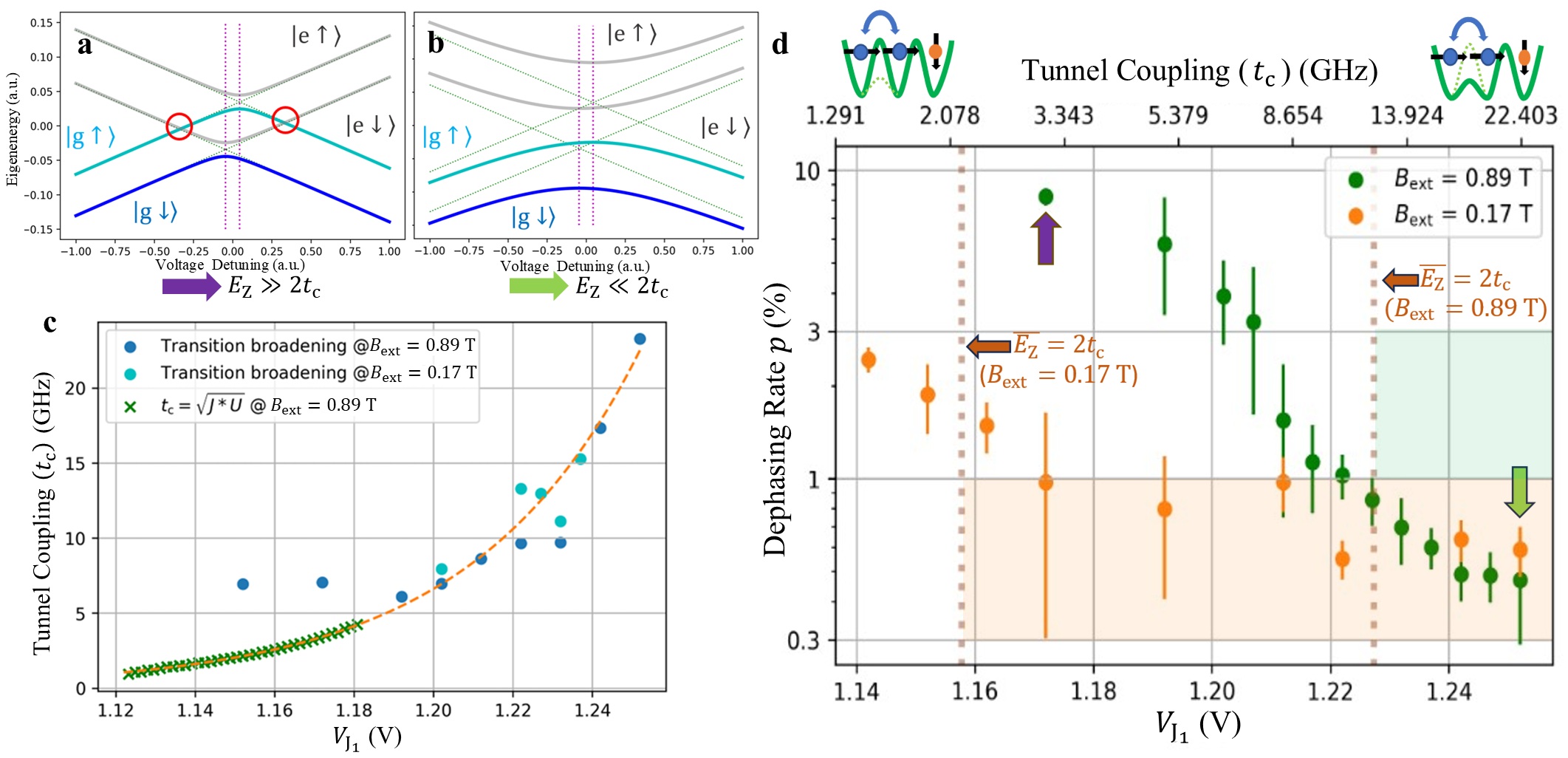} 
    
	\caption{\textbf{The tunnel-coupling dependence.}
        \textbf{a,} A schematic of the energy diagram when $\Ez \gg \ttc$; e.g., at the point indicated by the purple up arrow in (\textbf{d}). The eigenenergies of these four states are calculated from diagonalizing the Hamiltonian in Appendix C. The dotted magenta lines indicate the transitions between two charge states for spin-up (cyan) and spin-down (blue) states. Two degenerate points of the $\ket{\mathrm{g},\uparrow}$ (cyan) and $\ket{\mathrm{e}, \downarrow}$ (gray) states are indicated by red circles.
        \textbf{b,} A schematic of the energy diagram when $\Ez \ll \ttc$, e.g., the point indicated by the green-yellow down arrow in (\textbf{d}). In this case, the energy difference between spin-up and spin-down changes slowly compared to the case in whch $\Ez \gg \ttc$ in (\textbf{a}). Furthermore, the chemical potential of the orbital excited states is always higher than that of the ground states. 
        \textbf{c,} The tunnel couplings as a function of the gate voltage $\VJ$ are determined from both charge-transition broadening and spin exchange-coupling (see Appendix D)(see also Fig. S5 in the Supplemental Meterial). From the fits, we can calculate the tunnel couplings $\tc = (0.32 \pm 0.01)\exp[(23.56 \pm 0.3)(\VJ -1.072)]$ as a function of $\VJ$.
        \textbf{d,} The dephasing rate $p$ of the shuttling process as a function of the barrier gate voltage $\VJ$. The corresponding tunnel coupling ($\tc$) is calculated from fits of the experiment results (c). The tunnel couplings $\tc$ that equal half of the Zeeman splitting $\Ez$ under 0.17- and 0.89- \SI{}{\tesla} magnetic fields are indicated by dotted sienna lines. We characterize these dephasing rates by varying the number of shuttles (see Fig.~\ref{fig:LF} and Fig.~\ref{fig:HF} in the Supplemental Material).
    }
	\label{fig:fig3}
\end{figure*}

The spin polarization after shuttling is then evaluated (Fig.~\ref{fig:fig1}e and f). We prepare the spins in different eigenstates and shuttled back and forth between $\mathrm{Q_1}$ and $\mathrm{Q_2}$ repeatedly, with each cycle taking \SI{20}{\nano\second}. We measure them along four different projections on the Bloch sphere, made up of $x$, $y$, and the positive and negative $z$ directions. We will still require the operation of single-qubit gates to prepare both the spin-up and spin-down eigenstates as well as measure along the different projections as indicated (see Fig. \ref{fig:eigen} in the Supplemental Material). More details of this are given in Appendix B. Although the eigenstate shuttling error is small, it accumulates as the number of shuttles $N_\mathrm{sh}$ increases~\cite{Nat.Commun.12.4114}. We adjust the tunnel coupling between these two QDs by the gate voltage $\VJ$ to approximately $\tc\sim$ \SI{8.6}{\giga\Hz} (Fig.~\ref{fig:fig3}c), which is much larger than the Zeeman splittings $\Ez$ under the magnetic field of \SI{0.17}{\tesla}. From the fits, the depolarizing rates $r^{\downarrow(\uparrow)}$ for both spin-up and spin-down states are close to \SI{0.1}{\percent} for a return shuttling between $\mathrm{Q_1}$ and $\mathrm{Q_2}$. These error rates are considerably lower than the dephasing errors observed in the subsequent experiments, which is consistent with the expected error rates from charge shuttling.

We assess the phase coherence of shuttling using a Ramsey-type protocol (Fig.~\ref{fig:fig2}a-c). Initially, we apply a $X(\pi/2)$ gate to rotate the shuttling spin to a superposition state, while the ancilla spin remains in spin-down. Next, we move the spin from $\mathrm{Q_1}$ and $\mathrm{Q_2}$ by ramping the detuning of the quantum dots. We wait for a wait time $t_\mathrm{wait}$ in $\mathrm{Q_2}$, which is plotted on the $y$ axis in Fig.~\ref{fig:fig2}b. The spin is then shuttled back to $\mathrm{Q_1}$, and another $X(\pi/2)$ gate is applied to the shuttling spin for mapping the phase information to $Z$ projection. If the spin does cross the charge transition, the Larmor frequency of the shuttled spin will change, resulting in a noticeable difference in the precession rate. This transition in precession rate is clearly observed in Fig.~\ref{fig:fig2}b, demonstrating that the shuttling process is phase coherent. Furthermore, the difference in the precession rate of \SI{5.7}{\mega\Hz} aligns well with the resonance-frequency difference observed in the PESOS map (see Fig.~\ref{fig:CSD}c in the Supplemental Material.

To evaluate the dephasing error of the shuttling process, we prepare the spin in $\mathrm{Q_1}$ in an equal superposition state similar to what was done before and conduct consecutive shuttling between $\mathrm{Q_1}$ and $\mathrm{Q_2}$ (Fig.~\ref{fig:fig2}d-\ref{fig:fig2}f). To calculate the fidelity of the shuttling process, we model the shuttling process as a phase gate due to the spin precession at different Larmor frequencies in each dot, along with dephasing errors. Therefore, the possibility of measuring even states after $N_\mathrm{sh}$ shuttling operations can be expressed as $P_\mathrm{even} = \frac{A}{2}\exp\left(-(\kappa N_\mathrm{sh})^\beta\right)\left(1+\cos(\theta_\mathrm{sh} N_\mathrm{sh})\right)$. We fit $\theta_\mathrm{sh}$ and $\kappa$, which stands for the phase accumulated and the dephasing rate in one return shuttle between $\mathrm{Q_1}$ and $\mathrm{Q_2}$. The fitted $\theta_\mathrm{sh}$ remains stable with the number of shuttles, indicating the consistency of each shuttling operation, while the low decay rate $\kappa$ reflects the minimal dephasing rate $p = 1- \exp(-\kappa^\beta)$ = \SI{0.73}{\percent}. Considering shuttles as a phase gate along with both depolarizing and dephasing errors, we obtain the average shuttling fidelity over all pure input states to be $F_\mathrm{sh} = 1-\frac{1}{6}(r^\downarrow + r^\uparrow + 2p)$ = \SI{99.77}{\percent}~\cite{Nat.Commun.12.4114}.

\section{Tunnel Coupling dependence - Model}

We attribute the observed dephasing errors to two main sources: one occurring near the charge transitions of each spin state (dotted magenta lines in Fig.~\ref{fig:fig3}a) and the other at the degeneracies of any excited states(red circles in Fig.~\ref{fig:fig3}a). We describe our system by a $4 \times 4$ Hamiltonian consisting of both spin states and the orbital ground states in each QD (see Appendix C)~\cite{PhysRevB.107.085427}. When the voltage detuning $\varepsilon$ is far from the charge transition, the mixing of two charge states is negligible. In this scenario, the eigenstates for spin-up and spin-down have similar orbital states. Consequently, the energy difference between the spin-down and spin-up eigenstates is approximately equal to the Zeeman splitting in the dot. In contrast, as we approach $\varepsilon = 0$, the orbital state becomes a superposition of two charge states. Since the Zeeman splitting in each dot generally differs, which results from the different $g$ factors in our device, the charge-state transitions between two dots do not occur at the same detuning value for the spin-up and spin-down states (as indicated by the dashed magenta lines in Fig.~\ref{fig:fig3}a and \ref{fig:fig3}b). In the regime between these two transitions, the orbital states for the two spin states may vary significantly, affecting the effective $g$ factor. Therefore, the energy difference between the two spin states strongly depends on $\varepsilon$, making the phase coherence of the spin states sensitive to charge noise caused by fluctuations in the gate voltage.

To minimize this dephasing noise, it is important to ramp across both transitions as quickly as possible. However, the ramping rate is often constrained by the rise time of the wiring or the control hardware. Moreover, a faster ramp rate increases the likelihood of an LZ transition occurring at the transitions~\cite{PhysRevB.102.195418, j.physrep.2010.03.002, PhysRevB.102.125406}. Alternatively, this noise can be mitigated by increasing $\tc$, which allows for a smoother transition between charge states. When $\ttc \gg \dEz$, where $\dEz$ is the Zeeman-splitting difference between two dots, the orbital states of both spin-up and spin-down states resemble similar superposition states near the transitions. This similarity reduces the dependence of the energy difference on detuning, helping to alleviate the associated noise.

In addition to the noise origin from the different transitions, our model reveals two degeneracies between $\ket{\mathrm{e}\downarrow}$ and $\ket{\mathrm{g}\uparrow}$ states, where $\ket{\mathrm{e (g)}}$ stands for the orbital excited (ground) state, near the charge transitions (as indicated by the red circle in Fig.~\ref{fig:fig3}a). These additional degeneracies may cause spin-flipped tunneling due to the LZ transition~\cite{PhysRevB.102.125406}. In contrast, if $\ttc \gg \Ez$, the energy potentials of the $\ket{\mathrm{e}}$ are always higher than those of the $\ket{\mathrm{g}}$~(Fig.~\ref{fig:fig3}b)). In this scenario, no degeneracies between the orbital states occur across any regime. Furthermore, the significant anticrossing at the charge transition also reduces the likelihood of diabatic transits from ground states to excited states.

\section{Tunnel Coupling dependence - Experiment}

In this section, we focus on the dephasing error in the shuttling process between $\mathrm{Q_1}$ and $\mathrm{Q_2}$, between which the tunnel coupling can be controlled by the barrier gate voltage $\VJ$, and repeat consecutive-shuttling experiments from Fig.~\ref{fig:fig2}d-f. Instead of directly measuring the oscillation due to the phase precession, we construct the postshuttling states by measuring in various projections on the equator of the Bloch sphere (see Appendix B). This approach allows us to accurately ascertain the state, even after repeatedly shuttling and state decoherence. By analyzing the decoherence rate of the states as a function of the number of shuttling operations $N_\mathrm{sh}$, we determine the dephasing rate $p$.

We then measure the dephasing error rate at various gate voltages $\VJ$ while retaining the magnetic field at \SI{0.17}{\tesla}, which corresponds to a Zeeman splitting of $\Ez\sim$ \SI{4.76}{\giga\Hz}. Our findings show that the dephasing rate remains at approximately \SI{1}{\percent} until the voltage drops below \SI{1.16}{\volt}, where $\ttc \sim \Ez$. However, this decrease in fidelity might also arise from LZ transitions between two charge states at low tunnel coupling.

To address this, we increase the external magnetic field to \SI{0.89}{\tesla} to enhance the Zeeman splitting $\Ez$. Under this higher magnetic field, the condition $\Ez \ll \ttc$ is satisfied only at higher gate voltages $\VJ$ > \SI{1.23}{\volt}. The dephasing rate remains below \SI{1}{\percent} when this condition is satisfied. The difference in the error rates at high $\tc$, which leads to the average fidelity $F_\mathrm{sh} = $ \SI{99.81}{\percent}, is attributed to a difference in the $T_2^*$ of both QDs at the different magnetic fields (see Fig.~\ref{fig:T2} in the Supplemental Material). It also suggests that the remaining infidelity is limited by inherent dephasing over time rather than shuttling errors. Conversely, we observe an increase in the error rate when $\Ez \sim \ttc$, which occurs at higher $\VJ$ under this field strength. Therefore, we infer that this additional error stems from the error in spin shuttling, as previously mentioned.

\section{Conclusions}
In this work, we explore the impact of the tunnel coupling on the performance of the BB spin shuttling. We observe a reduction in phase error—nearly 20 times of magnitude—by adjusting the tunnel coupling. We attribute this enhancement in shuttling fidelity to the removal of decoherence hot spots associated with the spin states. By incorporating the spin states into our Hamiltonian, we highlight the degradation in performance that occurs when $\ttc \sim \Ez$; our experimental results accurately reflect this feature. At high tunnel coupling, we achieve shuttling fidelity up to \SI{99.8}{\percent} per shuttling in the MOS platform, though this is ultimately limited by the inherent dephasing time $T_2^*$. 

Additionally, we analyze the shuttling performance at a low magnetic field. To address the limitations imposed by the magnetic field strength~\cite{Nat.Commun.12.4114}, we replace Elzerman readout with PSB readout compared to the previous study~\cite{Nat.Commun.12.4114}. Unlike the significant change observed when sweeping the tunnel coupling under high field, the shuttling fidelity remains above \SI{99}{\percent} within the same range, which is consistent with our model. Further, the ability to do PSB readout also presents an opportunity for us to apply this technique in a scalable architecture, where we can reduce the fabrication overhead by reducing the need for reservoirs, as well as enable operations at high temperatures \cite{Nature.627.772-777}.

In conclusion, a large tunnel coupling is essential for maintaining phase coherence during spin shuttling. Our findings highlight the significance of accounting for the Zeeman splitting of the spin state alongside the conventional LZ transition of the charge state in the shuttling process. Operating under low magnetic fields is a straightforward approach to meet the requirement; besides, it also reduces the influence of Stark shift due to spin-orbital coupling when ramping the gate voltage. However, our results also manifest that the dephasing time is shorter, which we conjecture results from the lower magnetization of nuclear spins~\cite{Nat.Commun.10.5500, npj.Quantum.Inf.12.9}. Additionally, employing faster pulsing may also be crucial if the remaining error is linked to the inherent dephasing time. Investigating spin shuttling between quantum dots is vital for enabling nonlocal interaction in future large-scale systems or hopping-related techniques. The shuttling performance could be further enhanced by carefully balancing the shuttling parameters, and this optimization will be the subject of future work.

%% file: Content_2.tex
\section*{Acknowledgments}
We acknowledge support from the Australian Research Council (Grants No. FL190100167 and No. CE170100012), the U.S. Army Research Office (Grants No. W911NF-23-10092), the U.S. Air Force Office of Scientific Research (AFOSR) (Grants No. FA2386-22-1-4070), and the New South Wales Node of the Australian National Fabrication Facility. The views and conclusions contained in this document are those of the authors and should not be interpreted as representing the official policies, either expressed or implied, of the Army Research Office, the U.S. Air Force or the U.S. Government. The U.S. Government is authorised to reproduce and distribute reprints for Government purposes notwithstanding any copyright notation herein. S.C.L. acknowledges support from the National Science and Technology Council (NSTC) under Grants No. NSTC 112-2917-I-002-003. P.S. and A.D. acknowledge support from Sydney Quantum Academy. P.S. acknowledges support from the Baxter Charitable Foundation. H.S.G. acknowledges support from the NSTC, Taiwan, under Grants No. NSTC 113-2112-M-002-022-MY3, No. NSTC 113-2119-M-002-021, No. 114-2119-M-002-018, and No. NSTC 114-2119-M-002-017-MY3, from the AFOSR under Award No. FA2386-23-1-4052, and from the National Taiwan University under Grants No. NTU-CC-114L8950, No. NTU-CC114L895004, and No. NTU-CC-114L8517. H.S.G. is also grateful for the support from the "Center for Advanced Computing and Imaging in Biomedicine (Grant No. NTU-114L900702)" through the Featured Areas Research Center Program within the framework of the Higher Education Sprout Project by the Ministry of Education (MOE), Taiwan, from Taiwan Semiconductor Research Institute (TSRI) through the Joint Developed Project (JDP), and from the Physics Division, National Center for Theoretical Sciences, Taiwan. 

\section*{Author Contributions}
S.C.L., P.S., M.K.F., A.D., A.S., A.L., A.S.D., and C.H.Y. designed the experiments. S.C.L. and A.D. performed the experiments under the supervision of P.S., T.T., A.S.D, H.S.G., and C.H.Y.. W.H.L. and F.E.H. fabricated the device under A.S.D.’s supervision on enriched $^{28}$Si wafers supplied by K.M.I.. S.S. designed the RF-SET setup. P.S. and S.S. contributed to the experimental setup. M.K.F. and A.S. assisted in model construction and results analysis. S.C.L, P.S., M.K.F., and C.H.Y. wrote the manuscript, with input from all coauthors.

\section*{Corresponding Authors}
Correspondence to S.C.L., H.S.G, or C.H.Y..

\section*{Competing Interests}
A.S.D. is the CEO and a director of Diraq Pty Ltd. M.K.F, S.S, W.H.L., F.E.H., T.T., A.S., A.L., A.S.D., and C.H.Y. declare equity interest in Diraq Pty Ltd..


\section*{Data availability}
All data in this study are available from the Zenodo repository\cite{Zenodo.2507.15554}.


\section*{Appendices}

\section{Measurement Setup}
The device is measured in a K100 Kelvinox dilution refrigerator and mounted on the cold finger. An IPS120-10 Oxford superconducting magnet supplies the external magnetic field in the experiments. The magnetic field points along the [110] direction of the Si lattice. The voltages applied to the gates are produced by two sources and are combined via custom voltage combiners at room temperature. The QDevil QDAC supplies the dc voltages, while the Quantum Machine QPX+ generates the dynamic voltages for shuttling in a sample time of \SI{4}{\nano\second}. The dynamic pulse lines in the refrigerator have a bandwidth of 0$-$50 \SI{}{\mega\Hz}, which translates into a minimum rise time of \SI{20}{\nano\second}. A Keysight PSG8267D Vector Signal Generator mainly synthesizes the microwave pulses for the ESR single-quit control, utilizing the baseband IQ and pulse modulation signal from the OPX+. The modulated pulse can oscillate from \SI{250}{\kilo\Hz} up to \SI{44}{\giga\Hz}, although this range is also limited by the transmission line in the refrigerator.

The charge sensor consists of a single-island SET connected to a resonant $LC$ tank circuit on the sample PSB. A tone signal at \SI{165}{\mega\Hz} for the rf reflectometry is generated by the OPX+. The return signal is first amplified by a Cosmic Microwave Technology CITFL1 LNA at the 4-\SI{}{\kelvin} stage, and is subsequently amplified by two Mini-circuit ZFL-1000LN+ LNA devices at room temperature. After amplification, the signal is digitized and demodulated by the OPX+.

\section{Shuttling Characterization}

We begin by verifying the spin shuttling through the free precession, measuring along $X$, $Y$, and $Z$ projections by applying different single-qubit gates before the PSB readout (see Fig.~\ref{fig:XYZ} in the Supplemental Material). These single-qubit gates can be either $X(\pi)$ gates of $X(\pi/2)$ gates depending on the target projection. Similarly, $X(\pi)$ gates are also performed during initialization to prepare the shuttling spin in the spin-up state. 

\begin{figure}[htp!]
	\centering
	\includegraphics[width=0.4\textwidth]{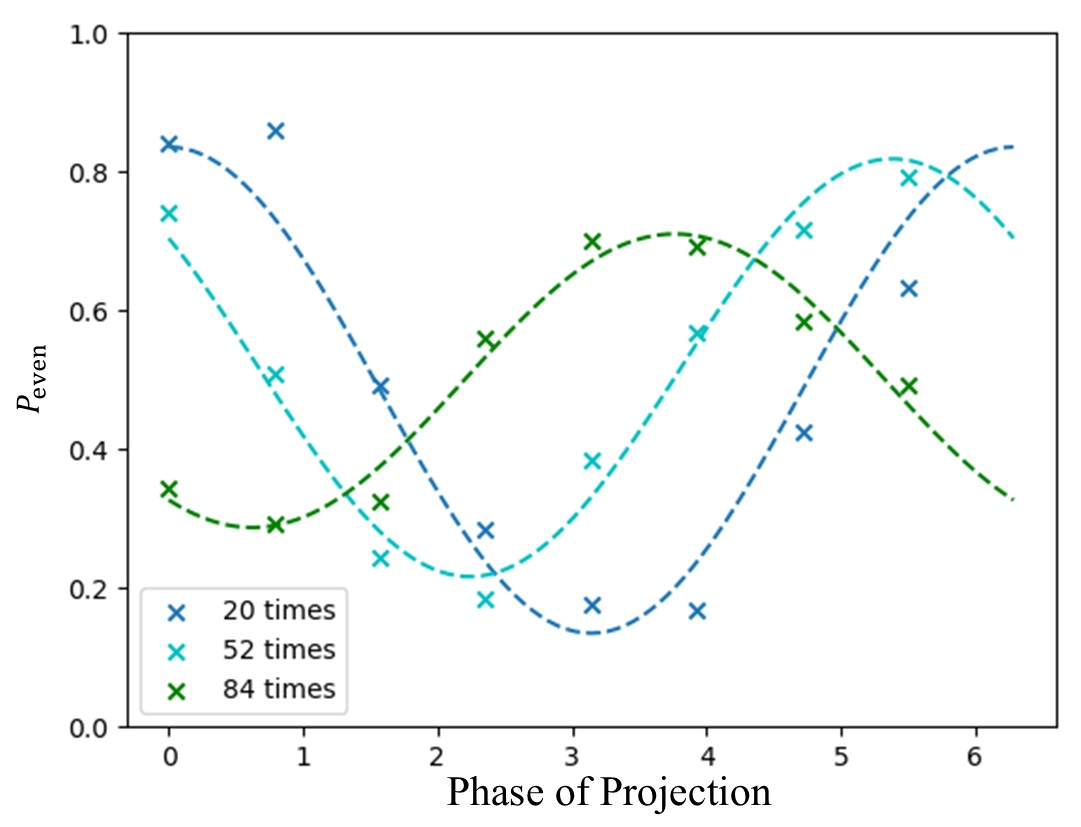} 
    
	\caption{\textbf{Spin-state assessment.}
        The postshuttling states are measured on various projections and the results are fitted by sinusoidal functions to determine the amplitude and the phase. The different-colored curves correspond to the projections measured after 20 (blue), 52 (cyan), and 84 (green) shuttling events, respectively.
    }
	\label{fig:phase}
\end{figure}

We appraise the error rate by the consecutive shuttling between $\mathrm{Q_1}$ and $\mathrm{Q_2}$. Each cycle of shuttling back and forth takes \SI{20}{\nano\second}, limited by the bandwidth of our experimental setup. We both prepare eigenstates to evaluate the depolarizing error and utilize equal superposition states on the equator of the Bloch sphere to assess the dephasing error. For the results of eigenstates, we observe a decay to one half for all three Pauli projections, with depolarization rates $r^{\downarrow(\uparrow)}$ close to \SI{0.1}{\percent}. In the case of shuttling equal superposition states, we measure the postshuttling states on various projections throughout the entire phase period on the equator to accurately determine the state~\cite{Nat.Commun.12.4114}. This is accomplished by applying an extra virtual phase gate before the second ESR $\frac{\pi}{2}$ gate before the readout. The resulting fringes are then fitted to sinusoidal functions (Fig.~\ref{fig:phase}), and the decay of the fringe amplitude as a function of the number of shuttles is calculated to estimate the dephasing rate $p$(see Fig.~\ref{fig:LF} and Fig.~\ref{fig:HF} in the Supplemental Material).

\section{Spin Shuttling Hamiltonian}
We describe our system by a Hamiltonian consisting of the Zeeman states $\{\ket{\uparrow}, \ket{\downarrow}\}$ and the lowest orbital states, $\{\ket{\mathrm{L}}, \ket{\mathrm{R}}\}$, in each QD, i.e.,  $\{\ket{\mathrm{L}, \uparrow},\ket{\mathrm{L}, \downarrow}, \ket{\mathrm{R}, \uparrow},\ket{\mathrm{R}, \downarrow}\}$. The Hamiltonian spanned under this basis is written as
\begin{widetext}
\begin{equation}
	H = \frac{1}{2}
        \begin{pmatrix}
            \varepsilon + g_\mathrm{L} \mu_\mathrm{B} B_\mathrm{ext} & 0 & \ttc & 0 \\
            0 & \varepsilon - g_\mathrm{L} \mu_\mathrm{B} B_\mathrm{ext} & 0 & \ttc \\
            \ttc & 0 &  -\varepsilon + g_\mathrm{R} \mu_\mathrm{B} B_\mathrm{ext} & 0 \\
            0 & \ttc & 0 & -\varepsilon - g_\mathrm{R} \mu_\mathrm{B} B_\mathrm{ext} \\
        \end{pmatrix},
	\label{eq:Hamiltonian}
\end{equation}
\end{widetext}
where $g_\mathrm{R}(g_\mathrm{L})$ is the $g$ factor in the right (left) dot. We can diagonalize this Hamiltonian and it results in $H = \frac{1}{2}\text{diag}(\Omega_- + \Ez, \Omega_+ - \Ez, -\Omega_- + \Ez, -\Omega_+ - \Ez)$, where $\Omega_\pm \equiv \sqrt{(\varepsilon\pm\frac{1}{2}\dEz)^2 + (\ttc)^2 }$, $\Ez \equiv \frac{1}{2}(g_\text{L} + g_\text{R})\mu_\mathrm{B} B_\mathrm{ext}$, and $\dEz \equiv (g_\text{R} - g_\text{L})\mu_\mathrm{B}  B_\mathrm{ext}$. The eigenergies as a function of the detuning $\varepsilon$ are plotted in Fig.~\ref{fig:fig3}a and \ref{fig:fig3}b.

Further, we can take the spin-dependent tunnel coupling $\tsd$ and spin-flip tunnel coupling $\tsf$ into account and focus on states $\ket{\Omega_+ - \bar{\Ez}}$ and $\ket{-\Omega_- + \Ez}$. Because $\tsd$ and $\tsf$ are typically much smaller than tunnel coupling $\tc$ \cite{PhysRevB.107.085427}, we can write down the perturbation term $V \sim 2\frac{\varepsilon\tsf  + \ttc\tsd}{\Omega_+ + \Omega_-} \ket{\Omega_+ - \Ez}\bra{-\Omega_- + \Ez} + \mathrm{h.c.}$ If $\ttc < \Ez$, this term may induce spin-flipped tunneling at $\Omega_\pm(\varepsilon) \approx \Ez$ (the indicated red circles in \ref{fig:fig3}a).

\section{Tunnel Coupling Assessment}
The tunnel couplings $\tc$ are first calculated from the broadening of the (101)-(011) interdots charge transition~\cite{PhysRevLett.92.226801}. Neglecting the spin freedom and describing the double QDs only by a two-level system consists of $\{\ket{\mathrm{L}}, \ket{\mathrm{R}}\}$. The eigenenergies are $\pm\frac{1}{2}\Omega $, and the corresponding eigenstates are $\ket{\mathrm{g(e)}}=\frac{1}{\sqrt{2\Omega(\Omega \mp \varepsilon)}}[\ttc\ket{\mathrm{L}} + (\pm\Omega - \varepsilon)\ket{\mathrm{R}}]$, where $\Omega \equiv \sqrt{\varepsilon^2+4{\tc}^2}$ and $\varepsilon = 0$ is set at the transition point. Assuming that the system is in thermodynamic equilibrium and follows the Fermi-Dirac distribution ${(\exp\frac{E_i-\mu}{k_\mathrm{B}T_\mathrm{e}} +  1)}^{-1}$, the average possibility of electron in the light (right) QD is $P = \frac{1}{2}(1 \mp \frac{\varepsilon}{\Omega}\tanh\frac{\Omega}{2k_\mathrm{B}T_\mathrm{e}})$ and it results in the SET signal
\begin{equation}
	I = I_0 + \frac{\partial I}{\partial\varepsilon}\varepsilon + \delta I\frac{\varepsilon}{\Omega}\tanh{\frac{\Omega}{2k_\mathrm{B}T_\mathrm{e}}}
	\label{eq:current}
\end{equation}
when sweeping the detuning. The first term stands for the background signal of the SET, while the second term accounts for the capacitance coupling of the gate voltages on the SET. Both of these terms are independent of the tunnel coupling and are excluded before fitting the measurement results. The results show the exponential dependence of the tunnel coupling on the gate voltage~\cite{PhysRevApplied.13.054018}. However, data points in the low-$\VJ$ regime exhibit deviations from the fitting, which is attributed to the resolution limit when sweeping the detuning, as well as to the constraints imposed by the finite electron temperature $T_\mathrm{e} \sim$ \SI{100}{\milli\kelvin}~\cite{PhysRevX.14.011048}.

We can calculate the tunnel coupling in the low-$\VJ$ regime from the exchange coupling $J$~\cite{PhysRevB.83.121403, Nano.Lett.2025.25.10263-10269} in the (110) charge state. The exchange couplings are characterized by the oscillation of decoupled controlled-phase gates (DCZs) at the symmetry point in the (110) charge state at different barrier gate voltages $\VJ$~\cite{Nature.601.343-347, Nat.Commun.16.3606}. The Hamiltonian in the basis ${\ket{\uparrow\uparrow}, \ket{\downarrow\uparrow}, \ket{\uparrow\downarrow}, \ket{\downarrow\downarrow}, \mathrm{S(2,0,0)}, \mathrm{S(0, 2, 0)}}$ is given by
\begin{equation}
	H = 
        \begin{pmatrix}
            \Ez & 0 & 0 & 0 & 0 & 0 \\
            0 & \frac{1}{2}\dEz & 0 & 0 & \frac{\tc}{\sqrt2} & \frac{\tc}{\sqrt2} \\
            0 & 0 &\frac{1}{2}\dEz & 0 & -\frac{\tc}{\sqrt2} & -\frac{\tc}{\sqrt2} \\
            0 & 0 & 0 & -\Ez & 0 & 0 \\
            0 & \frac{\tc}{\sqrt2} & -\frac{\tc}{\sqrt2} & 0 & U - \varepsilon & 0\\
            0 & \frac{\tc}{\sqrt2} & -\frac{\tc}{\sqrt2} & 0 & 0& U + \varepsilon \\
        \end{pmatrix},
    \label{eq:Hamiltonia_TcRaw}
\end{equation}
where $U$ is the charging energy to move both electrons into the same QD and $\varepsilon = 0$ is set at the symmetry point in (110) charge state. The charging energies $U$ are assumed to be the same for the S(200) and S(020) states. Hence, this equals the half of the voltage difference between the (200)-(110) and (110)-(020) charge transitions in the CSD. We separate diagonal and nondiagonal terms into $H_0$ and $V$. By choosing $S$ so that $[S,H_0] = -V$, 
\begin{equation}
	S = -\frac{\tc}{\sqrt2}
        \begin{pmatrix}
            0 & 0 & 0 & 0 & 0 & 0 \\
            0 & 0 & 0 & 0 & s_{--} & s_{+-} \\
            0 & 0 & 0 & 0 & -s_{-+} & -s_{++} \\
            0 & 0 & 0 & 0 & 0 & 0 \\
            0 & -s_{--} & s_{-+}  & 0 & 0 & 0\\
            0 & -s_{+-} & s_{++}  & 0 & 0 & 0 \\
        \end{pmatrix},
    \label{eq:Hamiltonian_Smatrix}
\end{equation}
where $s_{\pm\pm} \equiv (U\pm\varepsilon\pm\frac{1}{2}\dEz)^{-1}$. After the Schrieffer-Wolf transformation $e^S(H_0 + V)e^{-S} \approx H_0 + \frac{1}{2}[S, V]$, the effective Hamiltonian in the (110) subspace is
\begin{equation}
	H_{110} = 
        \begin{pmatrix}
            \Ez & 0 & 0 & 0 \\
            0 & \frac{1}{2}\dEz - J_- & J & 0 \\
            0 & J &-\frac{1}{2}\dEz  - J_+ & 0 \\
            0 & 0 & 0 & -\Ez \\
        \end{pmatrix},
	\label{eq:JCoupling}
\end{equation}
where $J_\pm \equiv \frac{\tc}{2}(s_{+\pm} + s_{-\pm})$ and $J \equiv \frac{1}{2}(J_+ + J_-)$. After diagonalization, this results in $H = \text{diag}(\Ez, \frac{1}{2}\Omega_J - J, -\frac{1}{2}\Omega_J - J, -\Ez)$, where $\Omega_J \equiv \sqrt{(\dEz + J_+ - J_-)^2 + 4J^2}$, under the basis ${\ket{\uparrow\uparrow}, \widetilde{\ket{\downarrow\uparrow}}, \widetilde{\ket{\uparrow\downarrow}}, \ket{\downarrow\downarrow}}$. If $\dEz \gg J$, the eigenstates $\widetilde{\ket{\downarrow\uparrow}} (\widetilde{\ket{\uparrow\downarrow}})$ are closed to the original $\ket{\downarrow\uparrow} (\ket{\uparrow\downarrow})$. In this case, the swap between $\ket{\downarrow\uparrow}$ and $\ket{\uparrow\downarrow}$ is negligible and a controlled-phase gate can be realized by the tuning exchange coupling $J$. Besides, We perform a dynamic decoupling pulse $\pi_{\mathrm{Q_1}}\pi_{\mathrm{Q_2}}$ in the middle of the DCZ gate wait time, which cancels the effect from $\Omega_J$. Because $U \gg \varepsilon,\dEz$, $J\approx J_\pm \approx\frac{\tc^2}{U}$.

%% file: Supplementary.tex
\section*{Supplementary Information}
\begin{figure*}[htp]
    \centering
	\includegraphics[width=1.0\textwidth]{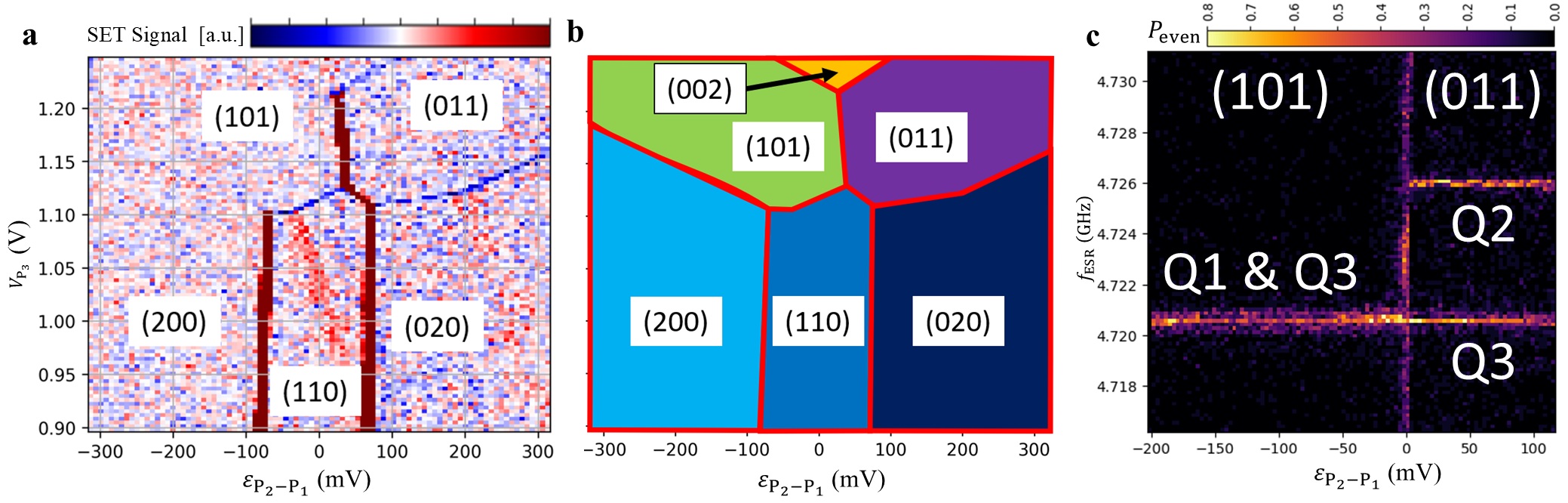} 
    
    \caption{\textbf{The charge-stability diagram (CSD) and pulsed electron spin-orbital
        spectroscopy (PESOS) Map.} 
        \textbf{a,} The original CSD with background noise. The dark red vertical lines correspond to the $\mathrm{Q_1 \text{-} Q_2}$ charge transitions. The tilted blue lines correspond to the $\mathrm{Q_2 \text{-} Q_3}$ charge transitions. The cotunnling event between (110) and (011) states are also shown in the plot. The two tilted blurred red transition in the (110) and (020) states correspond to the inter-dot orbital transition in $\mathrm{Q_2}$ because of the weak voltage confinement. By carefully tuning the voltage, we did not cross these transitions in our experiments. 
        \textbf{b,} A false-color CSD. The regime of each charge state is marked by one color and the transitions between them are indicated by the red lines. The transitions of $\mathrm{Q_1 \text{-} Q_3}$ cotunneling are not seen in our experiments (a) and the lines are only guess.
        \textbf{c,} The PESOS map near the (101)-(011) transition under an 0.17-\SI{}{\tesla} external field. The Larmor frequencies difference between the spins in $\mathrm{Q_1}$ and $\mathrm{Q_3}$ is too small under this external magnetic field strength, and hence their resonance frequencies almost overlap with each other at \SI{4.72}{\giga\Hz} in the plot.
    }
    \label{fig:CSD}
\end{figure*}

\begin{figure*}
    \centering
	\includegraphics[width=0.9\textwidth]{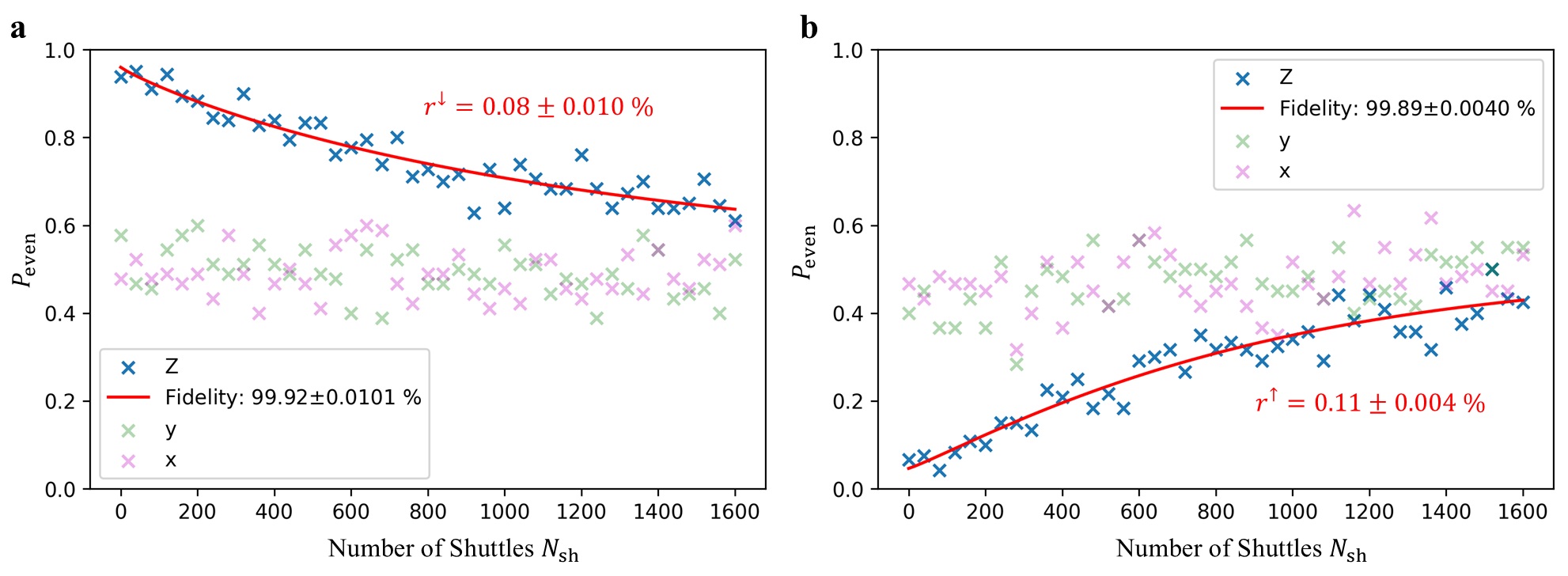} 

    \caption{\textbf{The polarization shuttling of the eigenstates.} \textbf{a,b,} In the eigenstates polarization shuttling, both spin-down (a) and spin-up (b) states are prepared and the post-shutting states are measured in all three Pauli basis of $\mathrm{Q_1}$. For the $Z$ projection, the states are measured on both positive $Z$ and negative $Z$ direction, by applying $\pi$ gate before measurement, to remove the background. The polarization amplitude of $Z- (-Z)$ results are not renormalized. The probabilities of finding spin-down (spin-up) after shuttling $N_\mathrm{sh}$ times is fitted with $P_{even} = \pm A\exp\left(-(\kappa_\uparrow N_\mathrm{sh})^\beta\right) + \frac{1}{2}$) and the depolarization rate $r^{\uparrow(\downarrow)} = 1 - \exp(\kappa^\beta)$.
    }
    \label{fig:eigen}
\end{figure*}

\pagebreak

\begin{figure*}
    \centering
	\includegraphics[width=0.9\textwidth]{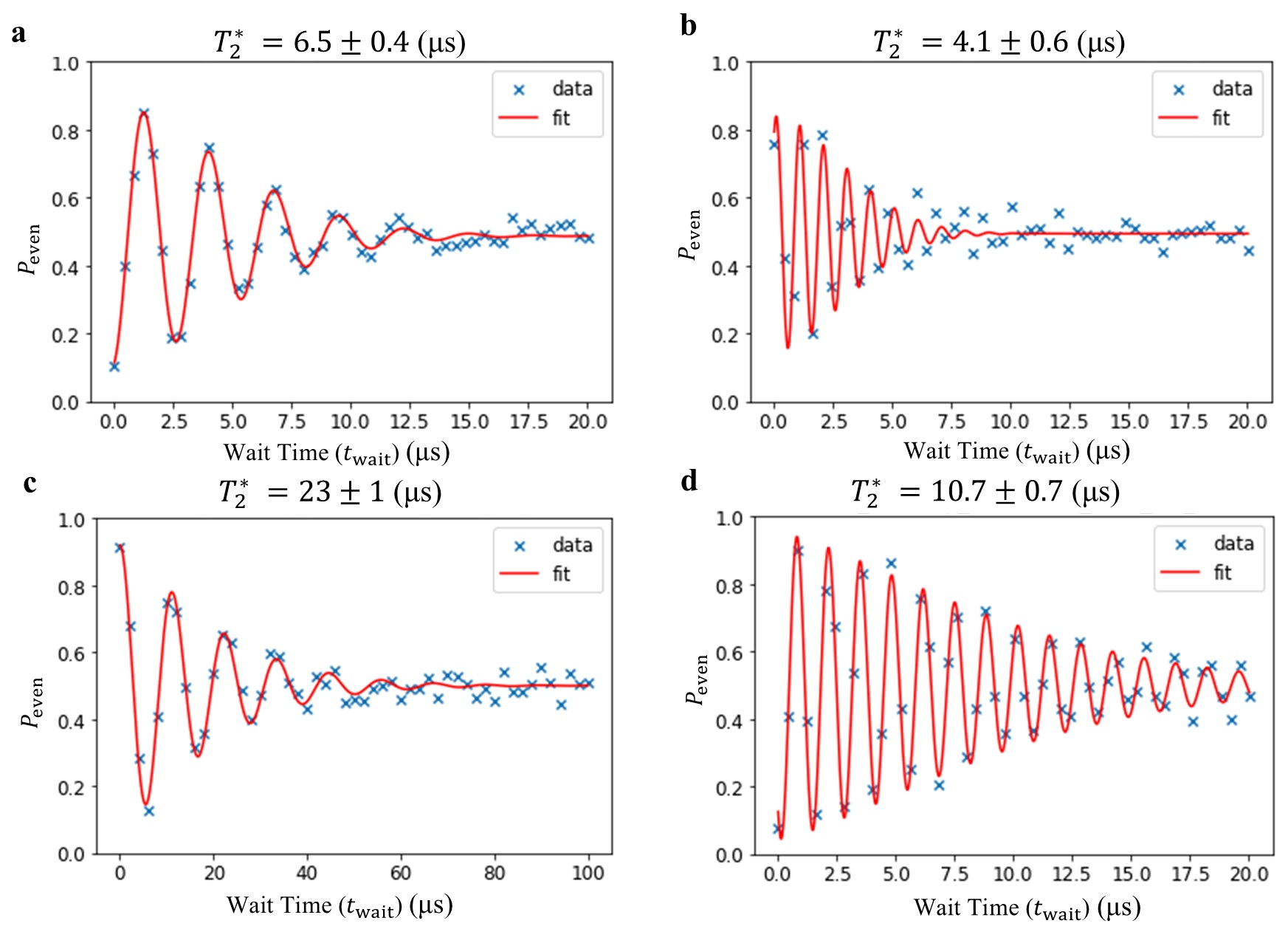} 

    \caption{\textbf{The dephasing time $T_2^*$.} The dephasing time $T_2^*$ of spin states are measured by Ramsey experiments at different charge states and under different magnetic fields. \textbf{a,} The spin in $\mathrm{Q_1}$ at (101) state under an 0.17-\SI{}{\tesla} external field. \textbf{b,} The spin in $\mathrm{Q_2}$ state at (011) state under an 0.17-\SI{}{\tesla} external field. \textbf{c,} The spin in $\mathrm{Q_1}$ at (101) state under an 0.89-\SI{}{\tesla} external field. \textbf{d,} The spin in $\mathrm{Q_2}$ at (011) state under an 0.89-\SI{}{\tesla} external field. The results are fitted with the function $P_\mathrm{even} = \frac{A}{2}\exp(-(t_\textrm{wait}/T_2^*)^\beta)(1+\cos\omega t_\textrm{wait})$. The higher $T_2^*$ under high external field is attributed to the better alignment of the residual nuclear spins.}
    \label{fig:T2}
\end{figure*}

\pagebreak

\begin{figure*}
    \centering
	\includegraphics[width=1.0\textwidth]{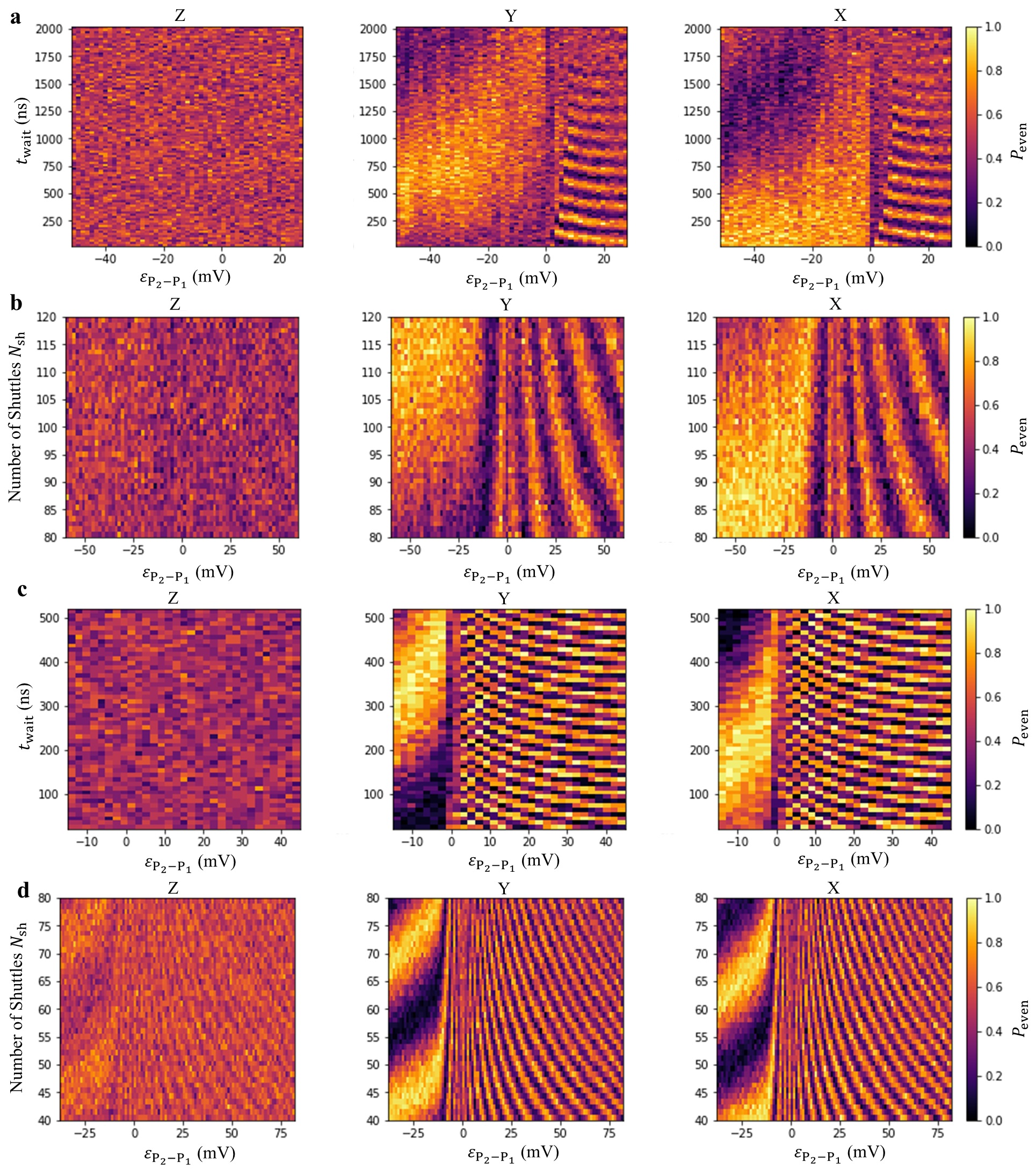} 

    \caption{\textbf{Different measurement projections.} The shuttling-spectroscopy measured in all three Pauli basis $X, Y$, and $Z$ projections of $\mathrm{Q_1}$. \textbf{a,b,} The shuttling spectroscopy (a) and consecutive-shuttling spectroscopy (b) under an 0.17-\SI{}{\tesla} external field. \textbf{c,d,} The shuttling spectroscopy (c) and consecutive-shuttling spectroscopy (d) under an 0.89-\SI{}{\tesla} external field.}
    \label{fig:XYZ}
\end{figure*}

\pagebreak

\begin{figure*}
    \centering
	\includegraphics[width=0.95\textwidth]{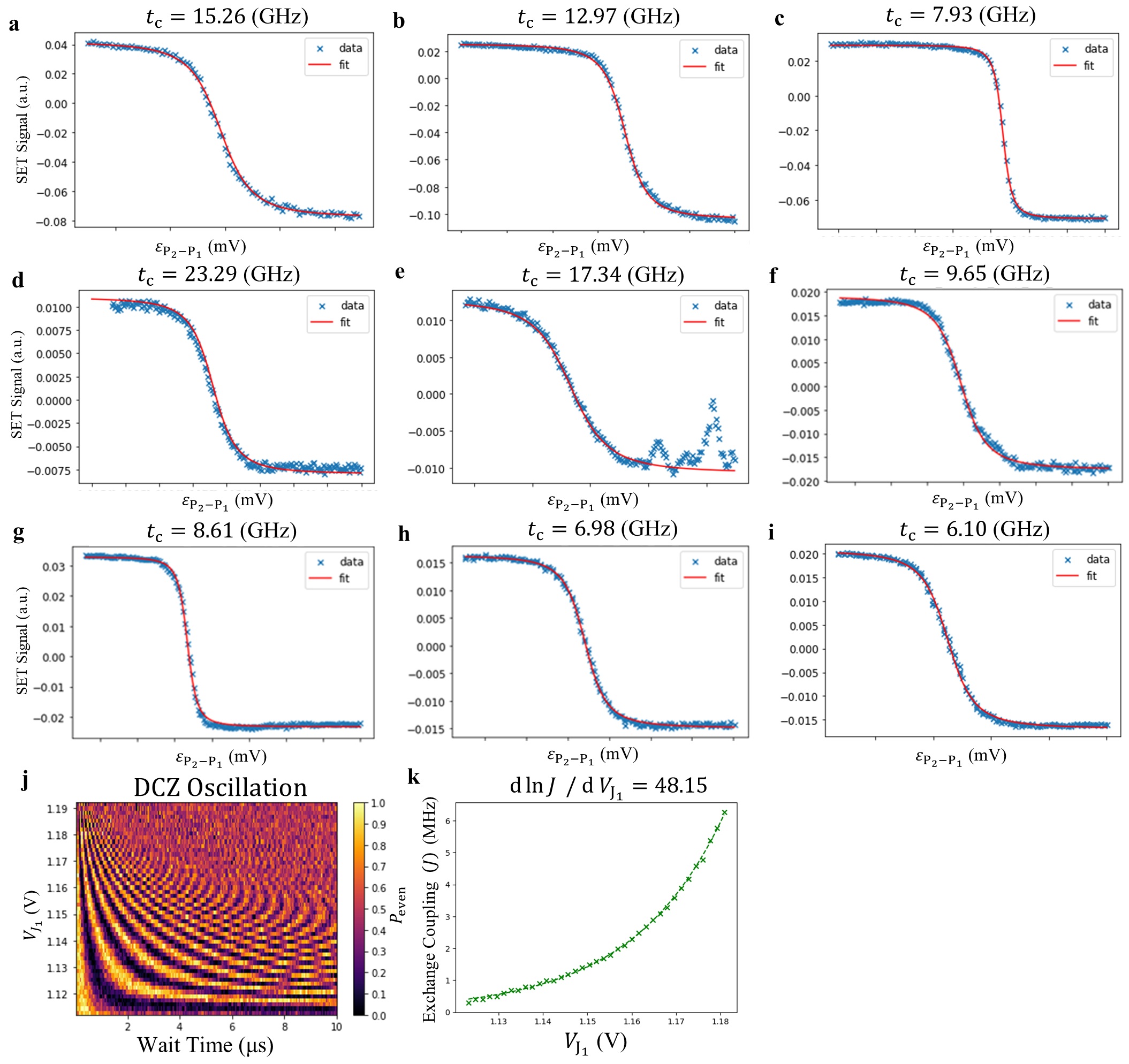} 
    
    \caption{\textbf{The tunnel couplings.} The tunnel coupling $\tc$ are calculated in two ways: charge-transition broadening or spin-exchange coupling. \textbf{a-c,} The charge-transition broadening between (101)-(011) under an 0.17-\SI{}{\tesla} external field at (a) $\VJ$ =\SI{ 1.237}{\volt}, (b) $\VJ$ = \SI{1.227}{\volt}, and (c) $\VJ$ = \SI{1.202}{\volt}. \textbf{d-i,} The charge transition broadening under an 0.89-\SI{}{\tesla} external field at (d) $\VJ$ = \SI{1.252}{\volt}, (e) $\VJ$ = \SI{1.242}{\volt}, (f) $\VJ$ = \SI{1.222}{\volt}, (g) $\VJ$ = \SI{1.212}{\volt}, (h) $\VJ$ = \SI{1.202}{\volt}, and (i) $\VJ$ = \SI{1.192}{\volt}. \textbf{j,} The oscillation of decoupled controlled-phase gates (DCZs) at (110) as a function of $\VJ$ and wait time. \textbf{k,} The spin-exchange coupling $J$ calculated from the Fourier transform of (j).}
    \label{fig:tc_detail}
\end{figure*}

\pagebreak

\begin{figure*}
    \centering
	\includegraphics[width=1.0\textwidth]{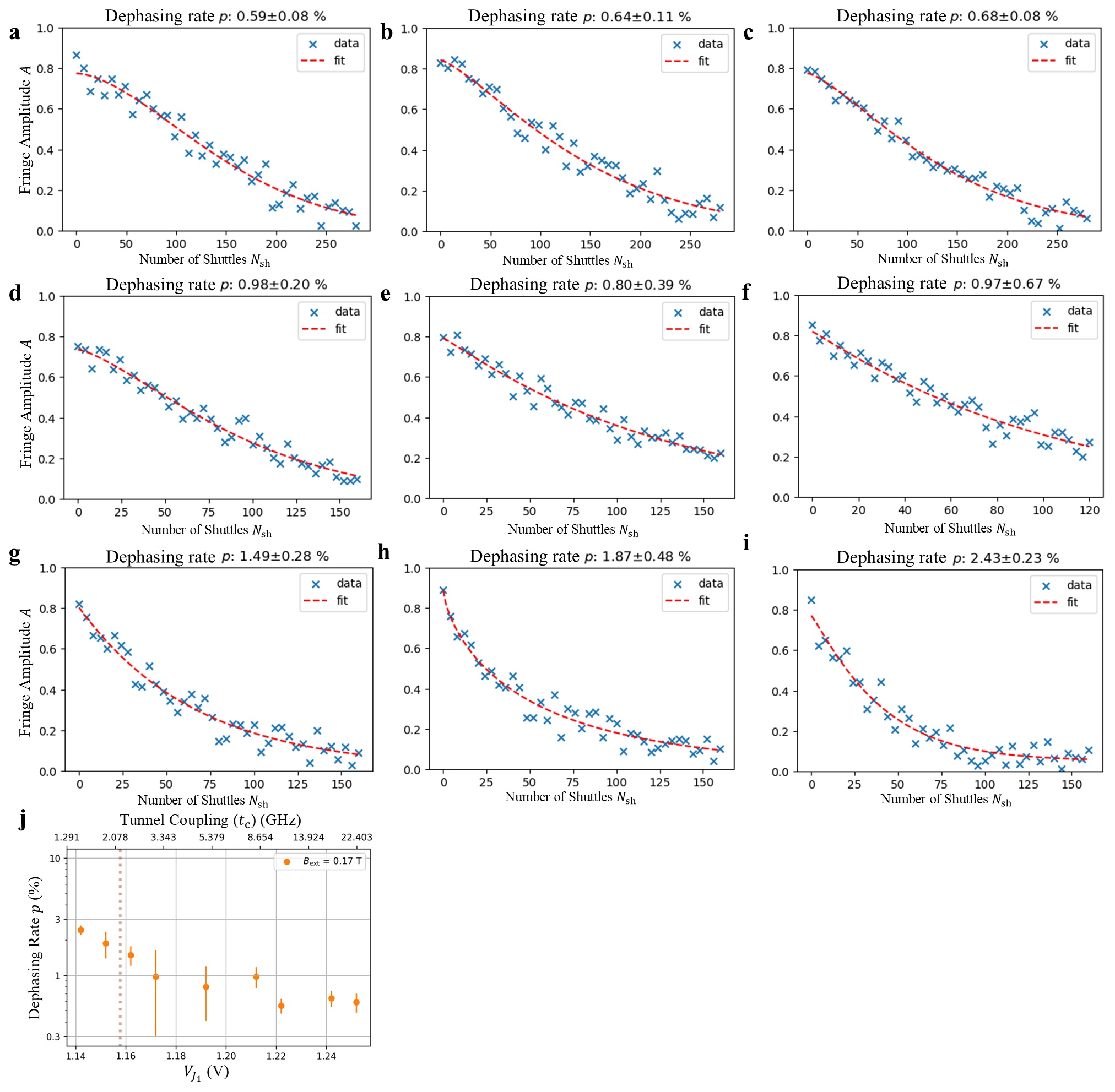} 

    \caption{\textbf{The consecutive shuttling at \SI{0.17}{\tesla}.} The fringe amplitude after consecutive shuttling as a function of the number of shuttles $N_\mathrm{sh}$ at (a) $V_\mathrm{J_1}$ = \SI{1.252}{\volt}, (b) $V_\mathrm{J_1}$ = \SI{1.242}{\volt}, (c) $V_\mathrm{J_1}$ = \SI{1.222}{\volt}, (d) $V_\mathrm{J_1}$ = \SI{1.212}{\volt}, (e) $V_\mathrm{J_1}$ = \SI{1.192}{\volt}, (f) $V_\mathrm{J_1}$ = \SI{1.172}{\volt}, (g) $V_\mathrm{J_1}$ = \SI{1.162}{\volt}, (h) $V_\mathrm{J_1}$ = \SI{1.152}{\volt}, and (i) $V_\mathrm{J_1}$ = \SI{1.142}{\volt}. The dashed red lines are the exponential fits of the decay $A = A_0 \exp(-(\kappa N_\mathrm{sh})^\beta) + d$ and the dephasing rates are calculated from $p = 1 - \exp(-\kappa^\beta)$. (j) Dephasing rate $p$ of the shuttling process as a function of barrier gate voltage $\VJ$. The tunnel coupling $\tc$ which equal half of the Zeeman splitting is indicated by sienna dotted lines.}
    \label{fig:LF}
\end{figure*}

\pagebreak

\begin{figure*}
    \centering
	\includegraphics[width=1.0\textwidth]{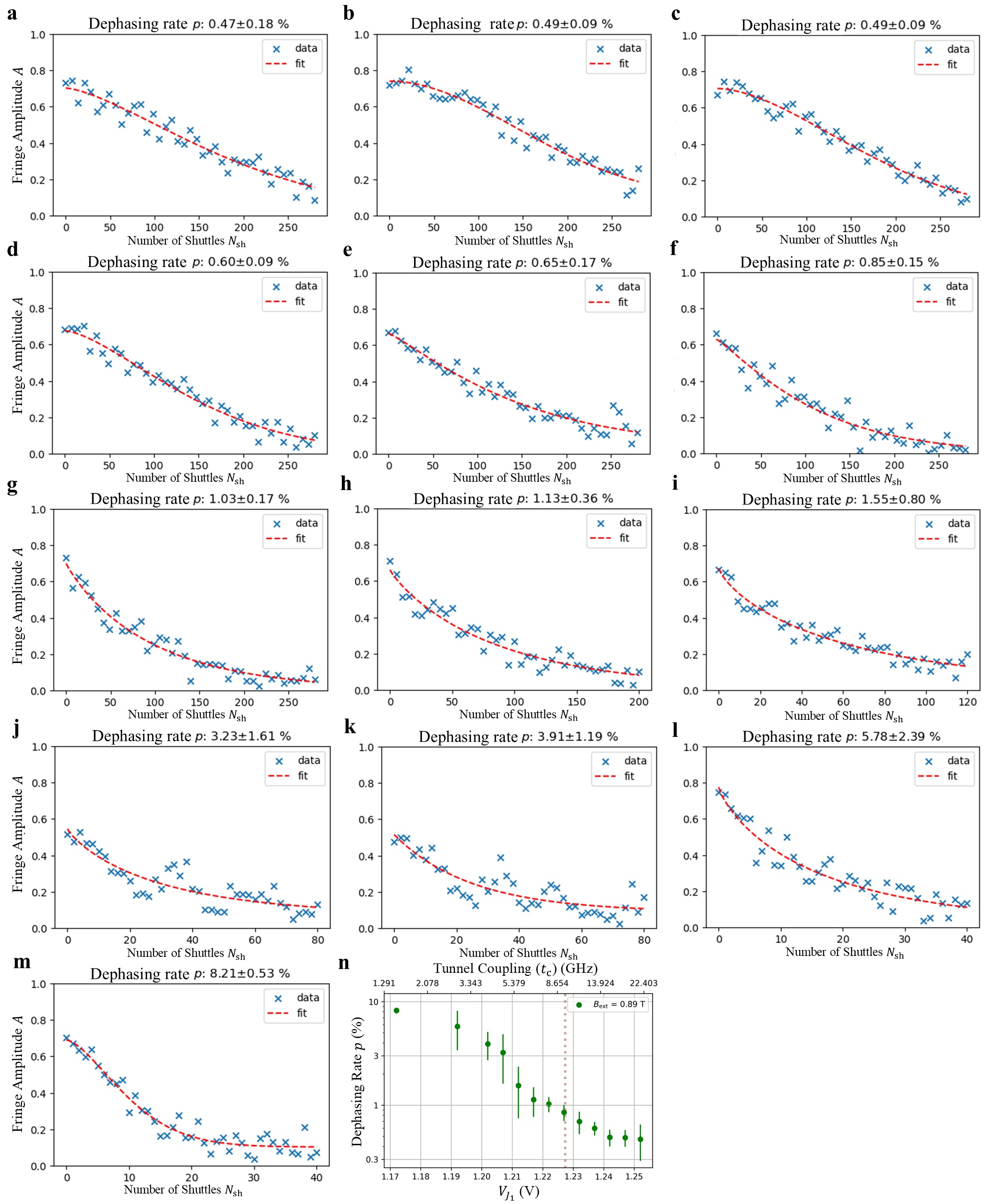} 

    \caption{\textbf{The consecutive shuttling at \SI{0.89}{\tesla}.} The fringe amplitude after consecutive shuttling as a function of the number of shuttles $N_\mathrm{sh}$ at (a) $V_\mathrm{J_1}$ = \SI{1.252}{\volt}, (b) $V_\mathrm{J_1}$ = \SI{1.247}{\volt}, (c) $V_\mathrm{J_1}$ = \SI{1.2442}{\volt}, (d) $V_\mathrm{J_1}$ = \SI{1.237}{\volt}, (e) $V_\mathrm{J_1}$ = \SI{1.232}{\volt}, (f) $V_\mathrm{J_1}$ = \SI{1.227}{\volt}, (g) $V_\mathrm{J_1}$ = \SI{1.222}{\volt}, (h) $V_\mathrm{J_1}$ = \SI{1.217}{\volt}, (i) $V_\mathrm{J_1}$ = \SI{1.212}{\volt}, (j) $V_\mathrm{J_1}$ = \SI{1.207}{\volt}, (k) $V_\mathrm{J_1}$ =\SI{ 1.202}{\volt}, (l) $V_\mathrm{J_1}$ = \SI{1.192}{\volt}, and (m) $V_\mathrm{J_1}$ = \SI{1.172}{\volt}. The dashed red lines are the fits and the the dephasing rates are calculated using the same formula as the ones under \SI{0.17}{\tesla}. (n) Dephasing rate $p$ of the shuttling process as a function of barrier gate voltage $\VJ$. The tunnel coupling $\tc$ which equal half of the Zeeman splitting is indicated by sienna dotted lines.}
    \label{fig:HF}
\end{figure*}

\pagebreak

%% file: bibliography.bib
@article{PRXQuantum.2.010303,
  title = {Pauli Blockade in Silicon Quantum Dots with Spin-Orbit Control},
  author = {Seedhouse, Amanda E. and Tanttu, Tuomo and Leon, Ross C.C. and Zhao, Ruichen and Tan, Kuan Yen and Hensen, Bas and Hudson, Fay E. and Itoh, Kohei M. and Yoneda, Jun and Yang, Chih Hwan and Morello, Andrea and Laucht, Arne and Coppersmith, Susan N. and Saraiva, Andre and Dzurak, Andrew S.},
  journal = {PRX Quantum},
  volume = {2},
  issue = {1},
  pages = {010303},
  numpages = {12},
  year = {2021},
  month = {Jan},
  publisher = {American Physical Society},
  doi = {10.1103/PRXQuantum.2.010303},
  url = {https://link.aps.org/doi/10.1103/PRXQuantum.2.010303}
}

@Article{Nat.Commun.12.4114,
    author={Yoneda, J. and Huang, W. and Feng, M. and Yang, C. H. and Chan, K. W. and Tanttu, T.
    and Gilbert, W. and Leon, R. C. C. and Hudson, F. E. and Itoh, K. M. and Morello, A.
    and Bartlett, S. D. and Laucht, A. and Saraiva, A. and Dzurak, A. S.},
    title={Coherent spin qubit transport in silicon},
    journal={Nat. Commun.},
    year={2021},
    month={Jul},
    day={05},
    volume={12},
    number={1},
    pages={4114},
    issn={2041-1723},
    doi={10.1038/s41467-021-24371-7},
    url={https://doi.org/10.1038/s41467-021-24371-7}
}

@article{PhysRevLett.92.226801,
  title = {Differential Charge Sensing and Charge Delocalization in a Tunable Double Quantum Dot},
  author = {DiCarlo, L. and Lynch, H. J. and Johnson, A. C. and Childress, L. I. and Crockett, K. and Marcus, C. M. and Hanson, M. P. and Gossard, A. C.},
  journal = {Phys. Rev. Lett.},
  volume = {92},
  issue = {22},
  pages = {226801},
  numpages = {4},
  year = {2004},
  month = {Jun},
  publisher = {American Physical Society},
  doi = {10.1103/PhysRevLett.92.226801},
  url = {https://link.aps.org/doi/10.1103/PhysRevLett.92.226801}
}

@Article{npj.Quantum.Inf.10.70,
    author={Elsayed, A. and Shehata, M. M. K. and Godfrin, C. and Kubicek, S. and Massar, S.
    and Canvel, Y. and Jussot, J. and Simion, G. and Mongillo, M. and Wan, D. and Govoreanu, B.
    and Radu, I. P. and Li, R. and Van Dorpe, P. and De Greve, K.},
    title={Low charge noise quantum dots with industrial CMOS manufacturing},
    journal={npj Quantum Inf.},
    year={2024},
    month={Jul},
    day={19},
    volume={10},
    number={1},
    pages={70},
    issn={2056-6387},
    doi={10.1038/s41534-024-00864-3},
    url={https://doi.org/10.1038/s41534-024-00864-3}
}

@Article{Nat.Electron.4.872-884,
    author={Gonzalez-Zalba, M. F. and de Franceschi, S. and Charbon, E. and Meunier, T.
    and Vinet, M. and Dzurak, A. S.},
    title={Scaling silicon-based quantum computing using CMOS technology},
    journal={Nat. Electron.},
    year={2021},
    month={Dec},
    day={01},
    volume={4},
    number={12},
    pages={872-884},
    issn={2520-1131},
    doi={10.1038/s41928-021-00681-y},
    url={https://doi.org/10.1038/s41928-021-00681-y}
}

@INPROCEEDINGS{IEEE.10185272,
  author={Elsayed, A. and Godfrin, C. and Dumoulin Stuyck, N.I. and Shehata, M.M.K. and Kubicek, S. and Massar, S. and Canvel, Y. and Jussot, J. and Hikavyy, A. and Loo, R. and Simion, G. and Mongillo, M. and Wan, D. and Govoreanu, B. and Li, R. and Radu, I.P. and Van Dorpe, P. and De Greve, K.},
  booktitle={2023 IEEE Symposium on VLSI Technology and Circuits (VLSI Technology and Circuits)}, 
  title={Comprehensive 300 mm process for Silicon spin qubits with modular integration}, 
  year={2023},
  volume={},
  number={},
  pages={1-2},
  doi={10.23919/VLSITechnologyandCir57934.2023.10185272},
  ISSN={2158-9682},
  month={June},
}

@Article{Nature.Nanotech.9.981-985,
    author={Veldhorst, M. and Hwang, J. C. C. and Yang, C. H. and Leenstra, A. W.
    and de Ronde, B. and Dehollain, J. P. and Muhonen, J. T. and Hudson, F. E.
    and Itoh, K. M. and Morello, A. and Dzurak, A. S.},
    title={An addressable quantum dot qubit with fault-tolerant control-fidelity},
    journal={Nat. Nanotechnol.},
    year={2014},
    month={Dec},
    day={01},
    volume={9},
    number={12},
    pages={981-985},
    issn={1748-3395},
    doi={10.1038/nnano.2014.216},
    url={https://doi.org/10.1038/nnano.2014.216}
}

@Article{Nature.601.343-347,
    author={Xue, Xiao and Russ, Maximilian and Samkharadze, Nodar and Undseth, Brennan
    and Sammak, Amir and Scappucci, Giordano and Vandersypen, Lieven M. K.},
    title={Quantum logic with spin qubits crossing the surface code threshold},
    journal={Nature (London)},
    year={2022},
    month={Jan},
    day={01},
    volume={601},
    number={7893},
    pages={343-347},
    issn={1476-4687},
    doi={10.1038/s41586-021-04273-w},
    url={https://doi.org/10.1038/s41586-021-04273-w}
}

@Article{Nature.601.338-342,
    author={Noiri, Akito and Takeda, Kenta and Nakajima, Takashi and Kobayashi, Takashi
    and Sammak, Amir and Scappucci, Giordano and Tarucha, Seigo},
    title={Fast universal quantum gate above the fault-tolerance threshold in silicon},
    journal={Nature (London)},
    year={2022},
    month={Jan},
    day={01},
    volume={601},
    number={7893},
    pages={338-342},
    issn={1476-4687},
    doi={10.1038/s41586-021-04182-y},
    url={https://doi.org/10.1038/s41586-021-04182-y}
    }

@article{Sci.Adv.8.eabn5130,
    author = {Adam R. Mills  and Charles R. Guinn  and Michael J. Gullans  and Anthony J. Sigillito  and Mayer M. Feldman  and Erik Nielsen  and Jason R. Petta },
    title = {Two-qubit silicon quantum processor with operation fidelity exceeding 99\%},
    journal = {Sci. Adv.},
    volume = {8},
    number = {14},
    pages = {eabn5130},
    year = {2022},
    doi = {10.1126/sciadv.abn5130},
    URL = {https://www.science.org/doi/abs/10.1126/sciadv.abn5130},
}

@Article{Nature.580.355-359,
    author={Petit, L. and Eenink, H. G. J. and Russ, M. and Lawrie, W. I. L.
    and Hendrickx, N. W. and Philips, S. G. J. and Clarke, J. S. and Vandersypen, L. M. K.
    and Veldhorst, M.},
    title={Universal quantum logic in hot silicon qubits},
    journal={Nature (London)},
    year={2020},
    month={Apr},
    day={01},
    volume={580},
    number={7803},
    pages={355-359},
    issn={1476-4687},
    doi={10.1038/s41586-020-2170-7},
    url={https://doi.org/10.1038/s41586-020-2170-7}
}

@Article{Nature.627.772-777,
    author={Huang, Jonathan Y. and Su, Rocky Y. and Lim, Wee Han and Feng, MengKe
    and van Straaten, Barnaby and Severin, Brandon and Gilbert, Will and Dumoulin Stuyck, Nard
    and Tanttu, Tuomo and Serrano, Santiago and Cifuentes, Jesus D. and Hansen, Ingvild
    and Seedhouse, Amanda E. and Vahapoglu, Ensar and Leon, Ross C. C. and Abrosimov, Nikolay V.
    and Pohl, Hans-Joachim and Thewalt, Michael L. W. and Hudson, Fay E. and Escott, Christopher C.
    and Ares, Natalia and Bartlett, Stephen D. and Morello, Andrea and Saraiva, Andre
    and Laucht, Arne and Dzurak, Andrew S. and Yang, Chih Hwan},
    title={High-fidelity spin qubit operation and algorithmic initialization above 1 K},
    journal={Nature (London)},
    year={2024},
    month={Mar},
    day={01},
    volume={627},
    number={8005},
    pages={772-777},
    issn={1476-4687},
    doi={10.1038/s41586-024-07160-2},
    url={https://doi.org/10.1038/s41586-024-07160-2}
}

@misc{arXiv.2409.03993,
      title={CMOS compatibility of semiconductor spin qubits}, 
      author={Nard Dumoulin Stuyck and Andre Saraiva and Will Gilbert and Jesus Cifuentes Pardo and Ruoyu Li and Christopher C. Escott and Kristiaan De Greve and Sorin Voinigescu and David J. Reilly and Andrew S. Dzurak},
      year={2024},
      eprint={2409.03993},
      archivePrefix={arXiv},
      primaryClass={cond-mat.mes-hall},
      url={https://arxiv.org/abs/2409.03993}, 
}

@Article{Nature.643.382-387,
    author={Bartee, Samuel K. and Gilbert, Will and Zuo, Kun and Das, Kushal and Tanttu, Tuomo
    and Yang, Chih Hwan and Dumoulin Stuyck, Nard and Pauka, Sebastian J. and Su, Rocky Y.
    and Lim, Wee Han and Serrano, Santiago and Escott, Christopher C. and Hudson, Fay E.
    and Itoh, Kohei M. and Laucht, Arne and Dzurak, Andrew S. and Reilly, David J.},
    title={Spin-qubit control with a milli-kelvin CMOS chip},
    journal={Nature (London)},
    year={2025},
    month={Jul},
    day={01},
    volume={643},
    number={8071},
    pages={382-387},
    issn={1476-4687},
    doi={10.1038/s41586-025-09157-x},
    url={https://doi.org/10.1038/s41586-025-09157-x}
}

@Article{Nat.Electron.4.64-70,
    author={Pauka, S. J. and Das, K. and Kalra, R. and Moini, A. and Yang, Y. and Trainer, M.
    and Bousquet, A. and Cantaloube, C. and Dick, N. and Gardner, G. C.
    and Manfra, M. J. and Reilly, D. J.},
    title={A cryogenic CMOS chip for generating control signals for multiple qubits},
    journal={Nat. Electron.},
    year={2021},
    month={Jan},
    day={01},
    volume={4},
    number={1},
    pages={64-70},
    issn={2520-1131},
    doi={10.1038/s41928-020-00528-y},
    url={https://doi.org/10.1038/s41928-020-00528-y}
}

@article{PhysRevApplied.18.024053,
  title = {Spiderweb Array: A Sparse Spin-Qubit Array},
  author = {Boter, Jelmer M. and Dehollain, Juan P. and van Dijk, Jeroen P.G. and Xu, Yuanxing and Hensgens, Toivo and Versluis, Richard and Naus, Henricus W.L. and Clarke, James S. and Veldhorst, Menno and Sebastiano, Fabio and Vandersypen, Lieven M.K.},
  journal = {Phys. Rev. Appl.},
  volume = {18},
  issue = {2},
  pages = {024053},
  numpages = {20},
  year = {2022},
  month = {Aug},
  publisher = {American Physical Society},
  doi = {10.1103/PhysRevApplied.18.024053},
  url = {https://link.aps.org/doi/10.1103/PhysRevApplied.18.024053}
}

@Article{Nat.Commun.15.4977,
    author={K{\"u}nne, Matthias and Willmes, Alexander and Oberl{\"a}nder, Max
    and Gorjaew, Christian and Teske, Julian D. and Bhardwaj, Harsh and Beer, Max
    and Kammerloher, Eugen and Otten, Ren{\'e} and Seidler, Inga and Xue, Ran
    and Schreiber, Lars R. and Bluhm, Hendrik},
    title={The SpinBus architecture for scaling spin qubits with electron shuttling},
    journal={Nat. Commun.},
    year={2024},
    month={Jun},
    day={11},
    volume={15},
    number={1},
    pages={4977},
    issn={2041-1723},
    doi={10.1038/s41467-024-49182-4},
    url={https://doi.org/10.1038/s41467-024-49182-4}
}

@Article{Nat.Commun.13.5740,
    author={Noiri, Akito and Takeda, Kenta and Nakajima, Takashi and Kobayashi, Takashi
    and Sammak, Amir and Scappucci, Giordano and Tarucha, Seigo},
    title={A shuttling-based two-qubit logic gate for linking distant silicon quantum processors},
    journal={Nat. Commun.},
    year={2022},
    month={Sep},
    day={30},
    volume={13},
    number={1},
    pages={5740},
    issn={2041-1723},
    doi={10.1038/s41467-022-33453-z},
    url={https://doi.org/10.1038/s41467-022-33453-z}
}

@Article{npj.Quantum.Inf.8.100,
    author={Seidler, Inga and Struck, Tom and Xue, Ran and Focke, Niels
    and Trellenkamp, Stefan and Bluhm, Hendrik and Schreiber, Lars R.},
    title={Conveyor-mode single-electron shuttling in Si/SiGe for a scalable quantum computing architecture},
    journal={npj Quantum Inf.},
    year={2022},
    month={Aug},
    day={30},
    volume={8},
    number={1},
    pages={100},
    issn={2056-6387},
    doi={10.1038/s41534-022-00615-2},
    url={https://doi.org/10.1038/s41534-022-00615-2}
}

@Article{Nat.Commun.15.2296,
    author={Xue, Ran and Beer, Max and Seidler, Inga and Humpohl, Simon and Tu, Jhih-Sian
    and Trellenkamp, Stefan and Struck, Tom and Bluhm, Hendrik and Schreiber, Lars R.},
    title={Si/SiGe QuBus for single electron information-processing devices with memory and micron-scale connectivity function},
    journal={Nat. Commun.},
    year={2024},
    month={Mar},
    day={14},
    volume={15},
    number={1},
    pages={2296},
    issn={2041-1723},
    doi={10.1038/s41467-024-46519-x},
    url={https://doi.org/10.1038/s41467-024-46519-x}
}

@article{PRXQuantum.5.040328,
  title = {Towards Early Fault Tolerance on a $2\ifmmode\times\else\texttimes\fi{}N$ Array of Qubits Equipped with Shuttling},
  author = {Siegel, Adam and Strikis, Armands and Fogarty, Michael},
  journal = {PRX Quantum},
  volume = {5},
  issue = {4},
  pages = {040328},
  numpages = {22},
  year = {2024},
  month = {Nov},
  publisher = {American Physical Society},
  doi = {10.1103/PRXQuantum.5.040328},
  url = {https://link.aps.org/doi/10.1103/PRXQuantum.5.040328}
}

@article{Appl.Phys.Lett.92.112103,
    author = {Angus, S. J. and Ferguson, A. J. and Dzurak, A. S. and Clark, R. G.},
    title = "{A silicon radio-frequency single electron transistor}",
    journal = {Appl. Phys. Lett.},
    volume = {92},
    number = {11},
    pages = {112103},
    year = {2008},
    month = {03},
    issn = {0003-6951},
    doi = {10.1063/1.2831664},
    url = {https://doi.org/10.1063/1.2831664},
}

@article{Appl.Phys.Rev.10.021305,
    author = {Vigneau, Florian and Fedele, Federico and Chatterjee, Anasua and Reilly, David and Kuemmeth, Ferdinand and Gonzalez-Zalba, M. Fernando and Laird, Edward and Ares, Natalia},
    title = "{Probing quantum devices with radio-frequency reflectometry}",
    journal = {Appl. Phys. Rev.},
    volume = {10},
    number = {2},
    pages = {021305},
    year = {2023},
    month = {02},
    issn = {1931-9401},
    doi = {10.1063/5.0088229},
    url = {https://doi.org/10.1063/5.0088229},
}

@Article{Nat.Electron.2.151-158,
    author={Yang, C. H. and Chan, K. W. and Harper, R. and Huang, W. and Evans, T.
    and Hwang, J. C. C. and Hensen, B. and Laucht, A. and Tanttu, T. and Hudson, F. E.
    and Flammia, S. T. and Itoh, K. M. and Morello, A. and Bartlett, S. D. and Dzurak, A. S.},
    title={Silicon qubit fidelities approaching incoherent noise limits via pulse engineering},
    journal={Nat. Electron.},
    year={2019},
    month={Apr},
    day={01},
    volume={2},
    number={4},
    pages={151-158},
    issn={2520-1131},
    doi={10.1038/s41928-019-0234-1},
    url={https://doi.org/10.1038/s41928-019-0234-1}
}

@Article{Nat.Commun.16.3606,
    author={Steinacker, Paul and Tanttu, Tuomo and Lim, Wee Han and Dumoulin Stuyck, Nard
    and Feng, MengKe and Serrano, Santiago and Vahapoglu, Ensar and Su, Rocky Y.
    and Huang, Jonathan Y. and Jones, Cameron and Itoh, Kohei M. and Hudson, Fay E.
    and Escott, Christopher C. and Morello, Andrea and Saraiva, Andre
    and Yang, Chih Hwan and Dzurak, Andrew S. and Laucht, Arne},
    title={Bell inequality violation in gate-defined quantum dots},
    journal={Nat. Commun.},
    year={2025},
    month={Apr},
    day={22},
    volume={16},
    number={1},
    pages={3606},
    issn={2041-1723},
    doi={10.1038/s41467-025-57987-0},
    url={https://doi.org/10.1038/s41467-025-57987-0}
}

@article{Appl.Phys.Lett.95.242102,
    author = {Lim, W. H. and Zwanenburg, F. A. and Huebl, H. and Möttönen, M. and Chan, K. W. and Morello, A. and Dzurak, A. S.},
    title = {Observation of the single-electron regime in a highly tunable silicon quantum dot},
    journal = {Appl. Phys. Lett.},
    volume = {95},
    number = {24},
    pages = {242102},
    year = {2009},
    month = {12},
    issn = {0003-6951},
    doi = {10.1063/1.3272858},
    url = {https://doi.org/10.1063/1.3272858},
}

@article{PhysRevB.107.085427,
  title = {Control of dephasing in spin qubits during coherent transport in silicon},
  author = {Feng, MengKe and Yoneda, Jun and Huang, Wister and Su, Yue and Tanttu, Tuomo and Yang, Chih Hwan and Cifuentes, Jesus D. and Chan, Kok Wai and Gilbert, William and Leon, Ross C. C. and Hudson, Fay E. and Itoh, Kohei M. and Laucht, Arne and Dzurak, Andrew S. and Saraiva, Andre},
  journal = {Phys. Rev. B},
  volume = {107},
  issue = {8},
  pages = {085427},
  numpages = {16},
  year = {2023},
  month = {Feb},
  publisher = {American Physical Society},
  doi = {10.1103/PhysRevB.107.085427},
  url = {https://link.aps.org/doi/10.1103/PhysRevB.107.085427}
}

@Article{Nature.646.81-87,
    author={Steinacker, Paul and Dumoulin Stuyck, Nard and Lim, Wee Han and Tanttu, Tuomo
    and Feng, MengKe and Serrano, Santiago and Nickl, Andreas and Candido, Marco
    and Cifuentes, Jesus D. and Vahapoglu, Ensar and Bartee, Samuel K. and Hudson, Fay E.
    and Chan, Kok Wai and Kubicek, Stefan and Jussot, Julien and Canvel, Yann
    and Beyne, Sofie and Shimura, Yosuke and Loo, Roger and Godfrin, Clement
    and Raes, Bart and Baudot, Sylvain and Wan, Danny and Laucht, Arne and Yang, Chih Hwan
    and Saraiva, Andre and Escott, Christopher C. and De Greve, Kristiaan and Dzurak, Andrew S.},
    title={Industry-compatible silicon spin-qubit unit cells exceeding 99{\%} fidelity},
    journal={Nature (London)},
    year={2025},
    month={Oct},
    day={01},
    volume={646},
    number={8083},
    pages={81-87},
    issn={1476-4687},
    doi={10.1038/s41586-025-09531-9},
    url={https://doi.org/10.1038/s41586-025-09531-9}
}

@article{PhysRevB.83.121403,
  title = {Efficient controlled-phase gate for single-spin qubits in quantum dots},
  author = {Meunier, T. and Calado, V. E. and Vandersypen, L. M. K.},
  journal = {Phys. Rev. B},
  volume = {83},
  issue = {12},
  pages = {121403},
  numpages = {4},
  year = {2011},
  month = {Mar},
  publisher = {American Physical Society},
  doi = {10.1103/PhysRevB.83.121403},
  url = {https://link.aps.org/doi/10.1103/PhysRevB.83.121403}
}

@article{Nano.Lett.2025.25.10263-10269,
    author = {Han Lim, Wee and Tanttu, Tuomo and Youn, Tony and Huang, Jonathan Yue and Serrano, Santiago and Dickie, Alexandra and Yianni, Steve and Hudson, Fay E. and Escott, Christopher C. and Yang, Chih Hwan and Laucht, Arne and Saraiva, Andre and Chan, Kok Wai and Cifuentes, Jesús D. and Dzurak, Andrew S.},
    title = {A 2 × 2 Quantum Dot Array in Silicon with Fully Tunable Pairwise Interdot Coupling},
    journal = {Nano Lett.},
    volume = {25},
    number = {26},
    pages = {10263-10269},
    year = {2025},
    doi = {10.1021/acs.nanolett.4c06264},
    note ={PMID: 40523105},
    URL = {https://doi.org/10.1021/acs.nanolett.4c06264}
}

@article{PRXQuantum.4.030303,
  title = {Shuttling an Electron Spin through a Silicon Quantum Dot Array},
  author = {Zwerver, A.M.J. and Amitonov, S.V. and de Snoo, S.L. and M\k{a}dzik, M.T. and Rimbach-Russ, M. and Sammak, A. and Scappucci, G. and Vandersypen, L.M.K.},
  journal = {PRX Quantum},
  volume = {4},
  issue = {3},
  pages = {030303},
  numpages = {11},
  year = {2023},
  month = {Jul},
  publisher = {American Physical Society},
  doi = {10.1103/PRXQuantum.4.030303},
  url = {https://link.aps.org/doi/10.1103/PRXQuantum.4.030303}
}

@article{PhysRevB.102.195418,
  title = {Spin shuttling in a silicon double quantum dot},
  author = {Ginzel, Florian and Mills, Adam R. and Petta, Jason R. and Burkard, Guido},
  journal = {Phys. Rev. B},
  volume = {102},
  issue = {19},
  pages = {195418},
  numpages = {13},
  year = {2020},
  month = {Nov},
  publisher = {American Physical Society},
  doi = {10.1103/PhysRevB.102.195418},
  url = {https://link.aps.org/doi/10.1103/PhysRevB.102.195418}
}

@article{PhysRevB.102.125406,
  title = {Simulated coherent electron shuttling in silicon quantum dots},
  author = {Buonacorsi, Brandon and Shaw, Benjamin and Baugh, Jonathan},
  journal = {Phys. Rev. B},
  volume = {102},
  issue = {12},
  pages = {125406},
  numpages = {11},
  year = {2020},
  month = {Sep},
  publisher = {American Physical Society},
  doi = {10.1103/PhysRevB.102.125406},
  url = {https://link.aps.org/doi/10.1103/PhysRevB.102.125406}
}

@article{PhysRevX.9.021028,
  title = {Controlling Spin-Orbit Interactions in Silicon Quantum Dots Using Magnetic Field Direction},
  author = {Tanttu, Tuomo and Hensen, Bas and Chan, Kok Wai and Yang, Chih Hwan and Huang, Wister Wei and Fogarty, Michael and Hudson, Fay and Itoh, Kohei and Culcer, Dimitrie and Laucht, Arne and Morello, Andrea and Dzurak, Andrew},
  journal = {Phys. Rev. X},
  volume = {9},
  issue = {2},
  pages = {021028},
  numpages = {10},
  year = {2019},
  month = {May},
  publisher = {American Physical Society},
  doi = {10.1103/PhysRevX.9.021028},
  url = {https://link.aps.org/doi/10.1103/PhysRevX.9.021028}
}

@article{PhysRevX.14.011048,
  title = {Exciton Transport in a Germanium Quantum Dot Ladder},
  author = {Hsiao, T.-K. and Cova Fari\~na, P. and Oosterhout, S. D. and Jirovec, D. and Zhang, X. and van Diepen, C. J. and Lawrie, W. I. L. and Wang, C.-A. and Sammak, A. and Scappucci, G. and Veldhorst, M. and Demler, E. and Vandersypen, L. M. K.},
  journal = {Phys. Rev. X},
  volume = {14},
  issue = {1},
  pages = {011048},
  numpages = {17},
  year = {2024},
  month = {Mar},
  publisher = {American Physical Society},
  doi = {10.1103/PhysRevX.14.011048},
  url = {https://link.aps.org/doi/10.1103/PhysRevX.14.011048}
}

@article{PRXQuantum.5.040322,
  title = {Strategies for Enhancing Spin-Shuttling Fidelities in $\mathrm{Si}$/$\mathrm{Si}$$\mathrm{Ge}$ Quantum Wells with Random-Alloy Disorder},
  author = {Losert, Merritt P. and Oberl\"ander, Max and Teske, Julian D. and Volmer, Mats and Schreiber, Lars R. and Bluhm, Hendrik and Coppersmith, S.N. and Friesen, Mark},
  journal = {PRX Quantum},
  volume = {5},
  issue = {4},
  pages = {040322},
  numpages = {26},
  year = {2024},
  month = {Nov},
  publisher = {American Physical Society},
  doi = {10.1103/PRXQuantum.5.040322},
  url = {https://link.aps.org/doi/10.1103/PRXQuantum.5.040322}
}

@Article{Nat.Commun.10.1063,
    author={Mills, A. R. and Zajac, D. M. and Gullans, M. J. and Schupp, F. J. and Hazard, T. M. 
            and Petta, J. R.},
    title={Shuttling a single charge across a one-dimensional array of silicon quantum dots},
    journal={Nat. Commun.},
    year={2019},
    month={Mar},
    day={05},
    volume={10},
    number={1},
    pages={1063},
    issn={2041-1723},
    doi={10.1038/s41467-019-08970-z},
    url={https://doi.org/10.1038/s41467-019-08970-z}
}

@article{Appl.Phys.Lett.124.114003,
    author = {Dumoulin Stuyck, Nard and Seedhouse, Amanda E. and Serrano, Santiago and Tanttu, Tuomo and Gilbert, Will and Huang, Jonathan Yue and Hudson, Fay and Itoh, Kohei M. and Laucht, Arne and Lim, Wee Han and Yang, Chih Hwan and Saraiva, Andre and Dzurak, Andrew S.},
    title = {Silicon spin qubit noise characterization using real-time feedback protocols and wavelet analysis},
    journal = {Appl. Phys. Lett.},
    volume = {124},
    number = {11},
    pages = {114003},
    year = {2024},
    month = {03},
    issn = {0003-6951},
    doi = {10.1063/5.0179958},
    url = {https://doi.org/10.1063/5.0179958},
}

@article{PhysRevB.92.201401,
  title = {Spin-orbit coupling and operation of multivalley spin qubits},
  author = {Veldhorst, M. and Ruskov, R. and Yang, C. H. and Hwang, J. C. C. and Hudson, F. E. and Flatt\'e, M. E. and Tahan, C. and Itoh, K. M. and Morello, A. and Dzurak, A. S.},
  journal = {Phys. Rev. B},
  volume = {92},
  issue = {20},
  pages = {201401},
  numpages = {5},
  year = {2015},
  month = {Nov},
  publisher = {American Physical Society},
  doi = {10.1103/PhysRevB.92.201401},
  url = {https://link.aps.org/doi/10.1103/PhysRevB.92.201401}
}

@Article{MRS.Communications.4.143-157,
    author={Itoh, Kohei M. and Watanabe, Hideyuki},
    title={Isotope engineering of silicon and diamond for quantum computing and sensing applications},
    journal={MRS Communications},
    year={2014},
    month={Dec},
    day={01},
    volume={4},
    number={4},
    pages={143-157},
    issn={2159-6867},
    doi={10.1557/mrc.2014.32},
    url={https://doi.org/10.1557/mrc.2014.32}
}

@Article{Nat.Commun.10.5500,
    author={Zhao, R. and Tanttu, T. and Tan, K. Y. and Hensen, B. and Chan, K. W. 
    and Hwang, J. C. C. and Leon, R. C. C. and Yang, C. H. and Gilbert, W. and Hudson, F. E.
    and Itoh, K. M. and Kiselev, A. A. and Ladd, T. D. and Morello, A. and Laucht, A.
    and Dzurak, A. S.},
    title={Single-spin qubits in isotopically enriched silicon at low magnetic field},
    journal={Nat. Commun.},
    year={2019},
    month={Dec},
    day={03},
    volume={10},
    number={1},
    pages={5500},
    issn={2041-1723},
    doi={10.1038/s41467-019-13416-7},
    url={https://doi.org/10.1038/s41467-019-13416-7}
}

@Article{Nat.Commun.16.5605,
    author={Unseld, Florian K. and Undseth, Brennan and Raymenants, Eline and Matsumoto, Yuta
    and de Snoo, Sander L. and Karwal, Saurabh and Pietx-Casas, Oriol and Ivlev, Alexander S.
    and Meyer, Marcel and Sammak, Amir and Veldhorst, Menno and Scappucci, Giordano
    and Vandersypen, Lieven M. K.},
    title={Baseband control of single-electron silicon spin qubits in two dimensions},
    journal={Nat. Commun.},
    year={2025},
    month={Jul},
    day={01},
    volume={16},
    number={1},
    pages={5605},
    issn={2041-1723},
    doi={10.1038/s41467-025-60351-x},
    url={https://doi.org/10.1038/s41467-025-60351-x}
}

@Article{npj.Quantum.Inf.10.22,
    author={Takeda, Kenta and Noiri, Akito and Nakajima, Takashi and Camenzind, Leon C.
    and Kobayashi, Takashi and Sammak, Amir and Scappucci, Giordano and Tarucha, Seigo},
    title={Rapid single-shot parity spin readout in a silicon double quantum dot with fidelity exceeding 99{\%}},
    journal={npj Quantum Inf.},
    year={2024},
    month={Feb},
    day={13},
    volume={10},
    number={1},
    pages={22},
    issn={2056-6387},
    doi={10.1038/s41534-024-00813-0},
    url={https://doi.org/10.1038/s41534-024-00813-0}
}

@article{Nano.Lett.2025.25.793−799,
    author = {George, Hubert C. and Mądzik, Mateusz T. and Henry, Eric M. and Wagner, Andrew J. and Islam, Mohammad M. and Borjans, Felix and Connors, Elliot J. and Corrigan, J. and Curry, Matthew and Harper, Michael K. and Keith, Daniel and Lampert, Lester and Luthi, Florian and Mohiyaddin, Fahd A. and Murcia, Sandra and Nair, Rohit and Nahm, Rambert and Nethwewala, Aditi and Neyens, Samuel and Patra, Bishnu and Raharjo, Roy D. and Rogan, Carly and Savytskyy, Rostyslav and Watson, Thomas F. and Ziegler, Josh and Zietz, Otto K. and Pellerano, Stefano and Pillarisetty, Ravi and Bishop, Nathaniel C. and Bojarski, Stephanie A. and Roberts, Jeanette and Clarke, James S.},
    title = {12-Spin-Qubit Arrays Fabricated on a 300 mm Semiconductor Manufacturing Line},
    journal = {Nano Lett.},
    volume = {25},
    number = {2},
    pages = {793-799},
    year = {2025},
    doi = {10.1021/acs.nanolett.4c05205},
    note ={PMID: 39721970},
    URL = {https://doi.org/10.1021/acs.nanolett.4c05205},
}

@Article{Nat.Phys.21.168-174,
    author={Dijkema, Jurgen and Xue, Xiao and Harvey-Collard, Patrick 
    and Rimbach-Russ, Maximilian and de Snoo, Sander L. and Zheng, Guojiand Sammak, Amir 
    and Scappucci, Giordano and Vandersypen, Lieven M. K.},
    title={Cavity-mediated iSWAP oscillations between distant spins},
    journal={Nat. Phys.},
    year={2025},
    month={Jan},
    day={01},
    volume={21},
    number={1},
    pages={168-174},
    issn={1745-2481},
    doi={10.1038/s41567-024-02694-8},
    url={https://doi.org/10.1038/s41567-024-02694-8}
}

@article{Microprocess.Microsyst.67.1-7,
    title = {Rent’s rule and extensibility in quantum computing},
    journal = {Microprocessors and Microsystems},
    volume = {67},
    pages = {1-7},
    year = {2019},
    issn = {0141-9331},
    doi = {https://doi.org/10.1016/j.micpro.2019.02.006},
    url = {https://www.sciencedirect.com/science/article/pii/S014193311830293X},
    author = {D.P. Franke and J.S. Clarke and L.M.K. Vandersypen and M. Veldhorst},
}

@article{qgnt-n527,
  title = {Mitigation of exchange crosstalk in dense quantum dot arrays},
  author = {Jirovec, Daniel and Fari\~na, Pablo Cova and Reale, Stefano and Oosterhout, Stefan D. and Zhang, Xin and de Snoo, Sander and Sammak, Amir and Scappucci, Giordano and Veldhorst, Menno and Vandersypen, Lieven M. K.},
  journal = {Phys. Rev. Appl.},
  volume = {24},
  issue = {3},
  pages = {034051},
  numpages = {23},
  year = {2025},
  month = {Sep},
  publisher = {American Physical Society},
  doi = {10.1103/qgnt-n527},
  url = {https://link.aps.org/doi/10.1103/qgnt-n527}
}

@article{xhq3-4jxz,
  title = {Operating Semiconductor Qubits without Individual Barrier Gates},
  author = {Ivlev, Alexander S. and Crielaard, Damien R. and Meyer, Marcel and Lawrie, William I. L. and Hendrickx, Nico W. and Sammak, Amir and Matsumoto, Yuta and Vandersypen, Lieven M. K. and Scappucci, Giordano and D\'eprez, Corentin and Veldhorst, Menno},
  journal = {Phys. Rev. X},
  volume = {15},
  issue = {3},
  pages = {031042},
  numpages = {10},
  year = {2025},
  month = {Aug},
  publisher = {American Physical Society},
  doi = {10.1103/xhq3-4jxz},
  url = {https://link.aps.org/doi/10.1103/xhq3-4jxz}
}

@Article{Nat.Nanotechnol.19.21-27,
author={Borsoi, Francesco and Hendrickx, Nico W. and John, Valentin and Meyer, Marcel
and Motz, Sayr and van Riggelen, Floor and Sammak, Amir and de Snoo, Sander L.
and Scappucci, Giordano and Veldhorst, Menno},
title={Shared control of a 16{\thinspace}semiconductor quantum dot crossbar array},
journal={Nat. Nanotechnol.},
year={2024},
month={Jan},
day={01},
volume={19},
number={1},
pages={21-27},
issn={1748-3395},
doi={10.1038/s41565-023-01491-3},
url={https://doi.org/10.1038/s41565-023-01491-3}
}

@misc{arXiv.2501.17814,
      title={A trilinear quantum dot architecture for semiconductor spin qubits}, 
      author={R. Li and V. Levajac and C. Godfrin and S. Kubicek and G. Simion and B. Raes and S. Beyne and I. Fattal and A. Loenders and W. De Roeck and M. Mongillo and D. Wan and K. De Greve},
      year={2025},
      eprint={2501.17814},
      archivePrefix={arXiv},
      primaryClass={quant-ph},
      url={https://arxiv.org/abs/2501.17814}, 
}

@article{Science309.2180-2184,
    author = {J. R. Petta  and A. C. Johnson  and J. M. Taylor  and E. A. Laird  and A. Yacoby  and M. D. Lukin  and C. M. Marcus  and M. P. Hanson  and A. C. Gossard },
    title = {Coherent Manipulation of Coupled Electron Spins in Semiconductor Quantum Dots},
    journal = {Science},
    volume = {309},
    number = {5744},
    pages = {2180-2184},
    year = {2005},
    doi = {10.1126/science.1116955},
    URL = {https://www.science.org/doi/abs/10.1126/science.1116955},
}

@Article{npj.Quantum.Inf.12.9,
    author={Rojas-Arias, Juan S. and Kojima, Yohei and Takeda, Kenta and Stano, Peter
    and Nakajima, Takashi and Yoneda, Junand Noiri, Akito and Kobayashi, Takashi
    and Loss, Daniel and Tarucha, Seigo},
    title={The origins of noise in the Zeeman splitting of spin qubits in natural-silicon devices},
    journal={npj Quantum Inf.},
    year={2025},
    month={Dec},
    day={09},
    volume={12},
    number={1},
    pages={9},
    issn={2056-6387},
    doi={10.1038/s41534-025-01150-6},
    url={https://doi.org/10.1038/s41534-025-01150-6}
}

@article{Adv.Mater.2023.35.2208557,
    author = {Wang, Zeheng and Feng, MengKe and Serrano, Santiago and Gilbert, William and Leon, Ross C. C. and Tanttu, Tuomo and Mai, Philip and Liang, Dylan and Huang, Jonathan Y. and Su, Yue and Lim, Wee Han and Hudson, Fay E. and Escott, Christopher C. and Morello, Andrea and Yang, Chih Hwan and Dzurak, Andrew S. and Saraiva, Andre and Laucht, Arne},
    title = {Jellybean Quantum Dots in Silicon for Qubit Coupling and On-Chip Quantum Chemistry},
    journal = {Adv. Mater.},
    volume = {35},
    number = {19},
    pages = {2208557},
    doi = {https://doi.org/10.1002/adma.202208557},
    url = {https://advanced.onlinelibrary.wiley.com/doi/abs/10.1002/adma.202208557},
    year = {2023}
}

@article{PhysRevB.110.125414,
  title = {Impact of electrostatic crosstalk on spin qubits in dense CMOS quantum dot arrays},
  author = {Cifuentes, Jesus D. and Tanttu, Tuomo and Steinacker, Paul and Serrano, Santiago and Hansen, Ingvild and Slack-Smith, James P. and Gilbert, Will and Huang, Jonathan Y. and Vahapoglu, Ensar and Leon, Ross C. C. and Stuyck, Nard Dumoulin and Itoh, Kohei and Abrosimov, Nikolay and Pohl, Hans-Joachim and Thewalt, Michael and Laucht, Arne and Yang, Chih Hwan and Escott, Christopher C. and Hudson, Fay E. and Lim, Wee Han and Rahman, Rajib and Dzurak, Andrew S. and Saraiva, Andre},
  journal = {Phys. Rev. B},
  volume = {110},
  issue = {12},
  pages = {125414},
  numpages = {8},
  year = {2024},
  month = {Sep},
  publisher = {American Physical Society},
  doi = {10.1103/PhysRevB.110.125414},
  url = {https://link.aps.org/doi/10.1103/PhysRevB.110.125414}
}

@Article{Nat.Nanotechnol.18.131-136,
    author={Gilbert, Will and Tanttu, Tuomo and Lim, Wee Han and Feng, MengKe
    and Huang, Jonathan Y. and Cifuentes, Jesus D. and Serrano, Santiago and Mai, Philip Y.
    and Leon, Ross C. C. and Escott, Christopher C. and Itoh, Kohei M. and Abrosimov, Nikolay V.
    and Pohl, Hans-Joachim and Thewalt, Michael L. W. and Hudson, Fay E. and Morello, Andrea
    and Laucht, Arne and Yang, Chih Hwan and Saraiva, Andre and Dzurak, Andrew S.},
    title={On-demand electrical control of spin qubits},
    journal={Nat. Nanotechnol.},
    year={2023},
    month={Feb},
    day={01},
    volume={18},
    number={2},
    pages={131-136},
    issn={1748-3395},
    doi={10.1038/s41565-022-01280-4},
    url={https://doi.org/10.1038/s41565-022-01280-4}
}

@article{PhysRevApplied.13.054018,
  title = {Efficient Orthogonal Control of Tunnel Couplings in a Quantum Dot Array},
  author = {Hsiao, T.-K. and van Diepen, C.J. and Mukhopadhyay, U. and Reichl, C. and Wegscheider, W. and Vandersypen, L.M.K.},
  journal = {Phys. Rev. Appl.},
  volume = {13},
  issue = {5},
  pages = {054018},
  numpages = {9},
  year = {2020},
  month = {May},
  publisher = {American Physical Society},
  doi = {10.1103/PhysRevApplied.13.054018},
  url = {https://link.aps.org/doi/10.1103/PhysRevApplied.13.054018}
}

@Article{Nat.Nanotechnol.20.866-872,
    author={De Smet, Maxim and Matsumoto, Yuta and Zwerver, Anne-Marije J. and Tryputen, Larysa
    and de Snoo, Sander L. and Amitonov, Sergey V. and Katiraee-Far, Sam R.
    and Sammak, Amir and Samkharadze, Nodar and G{\"u}l, {\"O}nder and Wasserman, Rick N. M.
    and Greplov{\'a}, Eli{\v{s}}ka and Rimbach-Russ, Maximilian and Scappucci, Giordano
    and Vandersypen, Lieven M. K.},
    title={High-fidelity single-spin shuttling in silicon},
    journal={Nat. Nanotechnol.},
    year={2025},
    month={Jul},
    day={01},
    volume={20},
    number={7},
    pages={866-872},
    issn={1748-3395},
    doi={10.1038/s41565-025-01920-5},
    url={https://doi.org/10.1038/s41565-025-01920-5}
}

@misc{arXiv.2506.15956,
      title={Scalable quantum current source on commercial 22-nm CMOS process technology}, 
      author={Ajit Dash and Suyash Pati Tripathi and Dimitrios Georgakopoulos and MengKe Feng and Steve Yianni and Ensar Vahapoglu and Md Mamunur Rahman and Shai Bonen and Owen Brace and Jonathan Y. Huang and Wee Han Lim and Kok Wai Chan and Will Gilbert and Arne Laucht and Andrea Morello and Andre Saraiva and Christopher C. Escott and Sorin P. Voinigescu and Andrew S. Dzurak and Tuomo Tanttu},
      year={2025},
      eprint={2506.15956},
      archivePrefix={arXiv},
      primaryClass={physics.app-ph},
      url={https://arxiv.org/abs/2506.15956}, 
}

@article{New.J.Phys.20.113029,
    doi = {10.1088/1367-2630/aaedd9},
    url = {https://dx.doi.org/10.1088/1367-2630/aaedd9},
    year = {2018},
    month = {nov},
    publisher = {IOP Publishing},
    volume = {20},
    number = {11},
    pages = {113029},
    author = {Li, Yi-Chao and Chen, Xi and Muga, J G and Sherman, E Ya},
    title = {Qubit gates with simultaneous transport in double quantum dots},
    journal = {New J. Phys.},
}

@article{j.physrep.2010.03.002,
    title = {Landau-Zener-Stückelberg interferometry},
    journal = {Physics Reports},
    volume = {492},
    number = {1},
    pages = {1-30},
    year = {2010},
    issn = {0370-1573},
    doi = {https://doi.org/10.1016/j.physrep.2010.03.002},
    url = {https://www.sciencedirect.com/science/article/pii/S0370157310000815},
    author = {S.N. Shevchenko and S. Ashhab and Franco Nori}
}

@dataset{Zenodo.2507.15554,
  author       = {Lin, Ssu-Chih and Steinacker, Paul and Feng, MengKe and
                  Dash, Ajit and Serrano, Santiago and Lim, Wee Han and
                  Itoh, Kohei M. and Hudson, Fay E. and Tanttu, Tuomo and
                  Saraiva, Andre and  Laucht, Arne and Dzurak, Andrew S. and
                  Goan, Hsi-Sheng and Yang, Chih Hwan},
  title        = {Data for "arXiv:2507.15554"},
  month        = aug,
  year         = 2025,
  publisher    = {Zenodo},
  doi          = {10.5281/zenodo.16927641},
  url          = {https://doi.org/10.5281/zenodo.16927641},
}
